\documentclass[12pt]{article}
\usepackage{epsf}
\usepackage{graphicx}
\usepackage{a4}
\usepackage{amsmath}
\usepackage{amssymb}
\usepackage{cite}		
\usepackage{color}
\usepackage{tikz}
\usepackage{pgfplots}
\usepackage{multicol}
\usepackage{multirow}
\usepackage{array}
\newcolumntype{P}[1]{>{\centering\arraybackslash}p{#1}}
\newcolumntype{M}[1]{>{\centering\arraybackslash}m{#1}}
\usepackage{empheq} 
\usepgfplotslibrary{fillbetween}

\def\hybrid{\topmargin 0pt
        \oddsidemargin 0pt
        \headheight 0pt \headsep 0pt
        \textwidth 6.25in       
        \textheight 9.5in       
        \marginparwidth .875in
        \parskip 5pt plus 1pt   \jot = 1.5ex}

\catcode`\@=11
\def\marginnote#1{}

\newcount\hour
\newcount\minute
\newtoks\amorpm
\hour=\time\divide\hour by60
\minute=\time{\multiply\hour by60 \global\advance\minute by-\hour}
\edef\standardtime{{\ifnum\hour<12 \global\amorpm={am}%
        \else\global\amorpm={pm}\advance\hour by-12 \fi
        \ifnum\hour=0 \hour=12 \fi
        \number\hour:\ifnum\minute<10 0\fi\number\minute\the\amorpm}}
\edef\militarytime{\number\hour:\ifnum\minute<10 0\fi\number\minute}

\def\draftlabel#1{{\@bsphack\if@filesw {\let\thepage\relax
   \xdef\@gtempa{\write\@auxout{\string
      \newlabel{#1}{{\@currentlabel}{\thepage}}}}}\@gtempa
   \if@nobreak \ifvmode\nobreak\fi\fi\fi\@esphack}
        \gdef\@eqnlabel{#1}}
\def\@eqnlabel{}
\def\@vacuum{}
\def\draftmarginnote#1{\marginpar{\raggedright\scriptsize\tt#1}}

\def\draft{\oddsidemargin -.5truein
        \def\@oddfoot{\sl preliminary draft \hfil
        \rm\thepage\hfil\sl\today\quad\militarytime}
        \let\@evenfoot\@oddfoot \overfullrule 3pt
        \let\label=\draftlabel
        \let\marginnote=\draftmarginnote
   \def\@eqnnum{(\theequation)\rlap{\kern\marginparsep\tt\@eqnlabel}%
\global\let\@eqnlabel\@vacuum}  }

\def\numberbysection{\@addtoreset{equation}{section}
        \def\theequation{\thesection.\arabic{equation}}}

\def\titlepage{\@restonecolfalse\if@twocolumn\@restonecoltrue\onecolumn
     \else \newpage \fi \thispagestyle{empty}\c@page\z@
        \def\thefootnote{\fnsymbol{footnote}}
	\setcounter{page}{0} }
\def\endtitlepage{\if@restonecol\twocolumn \else  \fi
        \def\thefootnote{\arabic{footnote}}
        \setcounter{footnote}{0}}  
\definecolor{c1}{rgb}{1, 0, 0}
\definecolor{c2}{rgb}{0, 1, 0}
\definecolor{c3}{rgb}{0, 0, 1}
\definecolor{c4}{rgb}{1, 0, 1}
\definecolor{c5}{rgb}{0, 1, 1}

\catcode`@=12
\relax

\def\beq{\begin{equation}}
\def\eeq{\end{equation}}
\def\bea{\begin{eqnarray}}
\def\eea{\end{eqnarray}}
\def\EQ{\begin{equation}}
\def\EN{\end{equation}}

\relax
\numberbysection
\hybrid

\begin{document}

\begin{center}
{\large\bf Long-range quenched bond disorder in the bi-dimensional Potts model}\\[.3in] 
{\bf Francesco\ Chippari}\\
  Sorbonne Universit\'e \& CNRS, UMR 7589, LPTHE, F-75005, Paris, France\\
    e-mail: {\tt fchippari@lpthe.jussieu.fr} \\
{\bf Marco\ Picco}\\
  Sorbonne Universit\'e \& CNRS, UMR 7589, LPTHE, F-75005, Paris, France\\
    e-mail: {\tt picco@lpthe.jussieu.fr} \\
    {\bf Raoul\ Santachiara}\\
  Paris-Saclay Universit\'e \& CNRS, UMR 8626, LPTMS, 91405, Saclay, France\\
    e-mail: {\tt raoul.santachiara@gmail.com}
\end{center}
\centerline{(Dated: \today)}
\vskip .2in
\centerline{\bf ABSTRACT}
\begin{quotation}
We study the bi-dimensional $q$-Potts model with long-range bond correlated disorder. Similarly to \cite{Chatelain}, we implement a disorder bimodal distribution by coupling the Potts model to auxiliary spin-variables, which are correlated with a power-law decaying function. 
The universal behaviour of different observables, especially the thermal and the order-parameter critical exponents, are computed  by Monte-Carlo techniques for $q=1,2,3$-Potts models for different values of the power-law decaying exponent $a$.
On the basis of our conclusions, which are in agreement with previous theoretical and numerical results for $q=1$ and $q=2$, we can conjecture the phase diagram for $q\in [1,4]$. In particular, we establish that the system is driven to a fixed point at finite or infinite  long-range disorder depending on the values of $q$ and $a$. 
Finally, we discuss the role of the higher cumulants of the disorder distribution. This is done by drawning the auxiliary spin-variables  from different statistical models. While the main features of the phase diagram depend only on the first and second cumulant, we argue, for the infinite disorder fixed point, that certain universal effects are affected by the higher cumulants of the disorder distribution.

\end{quotation}
\tableofcontents
\section{Introduction}
The $q$-Potts model \cite{wu1982potts} is a lattice spin model where the spins take one of $q$-possible values and interact via nearest neighbours couplings. Through a geometrical reformulation \cite{fortuin1972random}, the model admits a natural continuation to real values of $q$. When the couplings take the same positive value, the two-dimensional Potts model has a second-order ferromagnetic-paramagnetic phase transition for $q\in [1,4]$ \cite{wu1982potts}. Here, we consider the bond disordered $q$-Potts model, where the couplings have random fluctuations around a given positive value. In particular,  we are interested in how the critical properties of this model depend on the fluctuations (disorder) distribution. Part of the answer to this problem is provided by the Harris criterion \cite{harris1974effect} and by some generalizations of it \cite{WeinribHalperin,Weinrib}. These criteria determine the relevance of the disorder on the basis of the correlation properties of its distribution. The most studied example is the uncorrelated disorder where the couplings fluctuations are independent and equally distributed. In this case the Harris criterion states that the disorder modifies the critical properties of the pure system if its thermal exponent $\nu^{P}$ is greater than one,  $\nu^{P}>1$ \cite{harris1974effect}. This condition is satisfied by the $q$-Potts model for $q>2$. 
When the disorder is relevant, a situation of particular interest is the one where the system still undergoes a continuous phase transition but it is described by a different stable renormalization group (RG) fixed point. It is now well established that this is the situation for the two-dimensional Potts model with uncorrelated disorder. For $q>2$, there is a stable RG fixed point \cite{ludwig1990infinite,dotsenko1995renormalisation,picco1996numerical}, henceforth referred to as the Short-Range (SR) point, which determines the critical properties of the model. The thermal $\nu^{SR}$ and the order-parameter $\beta^{SR}$ critical exponents have been computed via an RG computation of a perturbed conformal field theory \cite{dotsenko1995renormalisation} and their values match with different numerical tests\cite{picco1996numerical,JacobsenCardy,chatelain99}.  
    
In this paper we consider the case where the couplings $\{J(x)\}$, at each position $x$, are drawn from an homogenous and isotropic distribution, with first cumulant, $\mathbb{E}[J(x)]=\mathbb{E}[J]$. We focus in particular on distributions whose second cumulant, $g(|x-y|)=\mathbb{E}[ J(x)J(y)]-\mathbb{E}[J]^2$, decreases as a power-law for large distances, $g(|x|)\propto (r-1)^2 |x|^{-a}$ for $|x|\gg1$. The parameter $a$ is taken positive, $a>0$. The value $r$ parametrizes the disorder strength and it is  taken greater than one, $r\in [1,\infty]$, see Section~\ref{numsetup}. For $r=1$, one recovers the pure Potts model, see Section~\ref{sec:r1}. The $r=\infty$ corresponds to the infinite disorder point and it is discussed in detail in Section~\ref{sec:rinfty}. The numerical set-up to implement these distributions follows closely the one introduced in \cite{Chatelain}, where the pure Potts model is coupled to the Ashkin-Teller one. Most of the results presented here are obtained by using a replicated Ising model instead, see Section~\ref{numsetup}, which differs from the one of \cite{Chatelain} in the cumulants of order higher than two. We have verified that our main findings on the phase diagram of these models do not depend on the higher cumulants of these distributions. 

The extended Harris criterion \cite{WeinribHalperin,Weinrib} determines in which region of the parameters $(q,a)$ the disorder is relevant and when its long-range 
properties prevail over the short-range ones. When $a>2$ the critical behaviour of the system is expected to be the same as the one with uncorrelated disorder. 
The regions ($q<2$, $a<2/\nu^{P}$) and ($q>2$, $a<2/\nu^{SR}$), where the disorder is relevant and dominated by the long-range correlations, are much less 
understood, especially for general values of $q$. 

For $q=1$, 
as we explain in Section~\ref{sec:q1}, the pure Potts model and the Potts with bimodal short range disorder correspond, 
for any value of the disorder strength $r$, to the Bernoulli bond percolation model \cite{SA92}, where the edges are independently activated (pure percolation). The long-range disordered $q=1$-Potts model coincides instead with a one-parameter (the disorder strength $r$) family of long-range bond percolation model, where the activation of two distant edges is correlated by a power-law decreasing function with exponent $a$.  The $d$-dimensional long-range percolation was studied in \cite{Weinrib} 
by using a one-loop order RG computation with a double expansion in $\epsilon$, $\epsilon= 6-d$ and in $\delta$, $\delta=4-a$. It was found that the long-range 
nature of the correlation drives the system to a new stable RG point, that, in the following,  we refer to as the LRp (Long-Range percolation) point. Even if these 
theoretical predictions have validity near dimension six and for $a$ close to $4$, they have motivated a series of numerical works \cite{stanley92, Schmittbuhl_1993,Janke_2017,de_Castro_2018,Javerzat_2020} on some $d=2$ long-range percolation models with $a>0$. These models were defined  by using 
the level sets of fractional Gaussian free fields with Hurst exponent $H=-a/2$. As the thermal exponent of the pure $q=1$-Potts is $\nu^{P}=4/3$, the long-range 
disorder is relevant for $a< 3/2$. The general agreement in \cite{stanley92, Schmittbuhl_1993,Janke_2017, de_Castro_2018,Javerzat_2020} was that, for $a<3/2$, 
the percolation transition is described by a new LRp point. In our approach, we recover the critical behaviour studied in \cite{stanley92, Schmittbuhl_1993,Janke_2017, de_Castro_2018,Javerzat_2020}  at the $r=\infty$ point of a Potts model coupled with a fractional Gaussian field. We will show that, irrespective to the value of $r$, 
these long-range percolation models are described, for $a>3/2$, by the Bernoulli percolation (Bp) point and, for $a<3/2$, by the LRp point.
This can be rephrased by saying that, for each $a$, there exists an unique fixed point which is stable in the direction of decreasing $r$. We observed that, by coupling instead the Potts model to $n-$replicated Ising models, the numerical investigations of the LRp type points becomes more precise, in particular for small values of $a$.  We will argue 
in Section~\ref{sec:hc} that some universal properties of the LRp points can depend on the higher cumulants of the disorder distribution.

The $q=2$ case corresponds to the bond disordered Ising model and it has been studied quite intensively in the past. Here the cross-over exponent is $2/\nu^{P}=2$. For $a\geq 2$, the SR and the pure (P) point coincide and the disorder effects generate logarithmic corrections to the scaling relations \cite{Dotsenko_1982}. For $1\lesssim a<2$, one finds a stable long-range (LR) point, whose existence has been proven in \cite{Rajabpour_2007} and in \cite{dudka2016critical} by using the Ising massive free fermion representation. Numerical tests of these predictions  have been found in \cite{Bagamery_2005} for $a=1$. 
In \cite{dudka2016critical} it was also observed that LR point looses its stability for $a\lesssim 1$,
and the question whether the system undergoes a smeared phase transition or whether it flows to an infinite disorder fixed point remained open. In \cite{Chatelain,Chatelain17}, the $q=2$ (and $q=4$) Potts long-range disordered model were investigated for values of $a$ smaller than one,  $0<a<1$. It was observed there that, for all this values of $a$, the Monte-Carlo measures of $\beta/\nu$ were very close to the LRp point value. 

Our aim here is to provide a complete Potts phase diagram in the parametric region $(q,a)\in (0,4]\times (0,\infty)$. This is achieved by considering the disordered $q=1,2,3$-Potts models for different values of $a$. These three values of $q$ are indeed  representative of the three regions where the short-range disorder is respectively irrelevant, marginal and relevant. 

Our results are summarised in Fig.~(\ref{ExtHarrfig}). We establish that for all values of $q>1$  there is a region where the LR point exists and it is stable in the direction of decreasing $r$. Moreover, for each $q\in [1,4]$, there is a value $a^{*}(q)$ at which the LR point becomes unstable. For smaller values of $a$, $a<a^{*}(q)$, the model is described by the LRp point, which is the fixed point of the Potts model with infinite disorder. 
\begin{center}
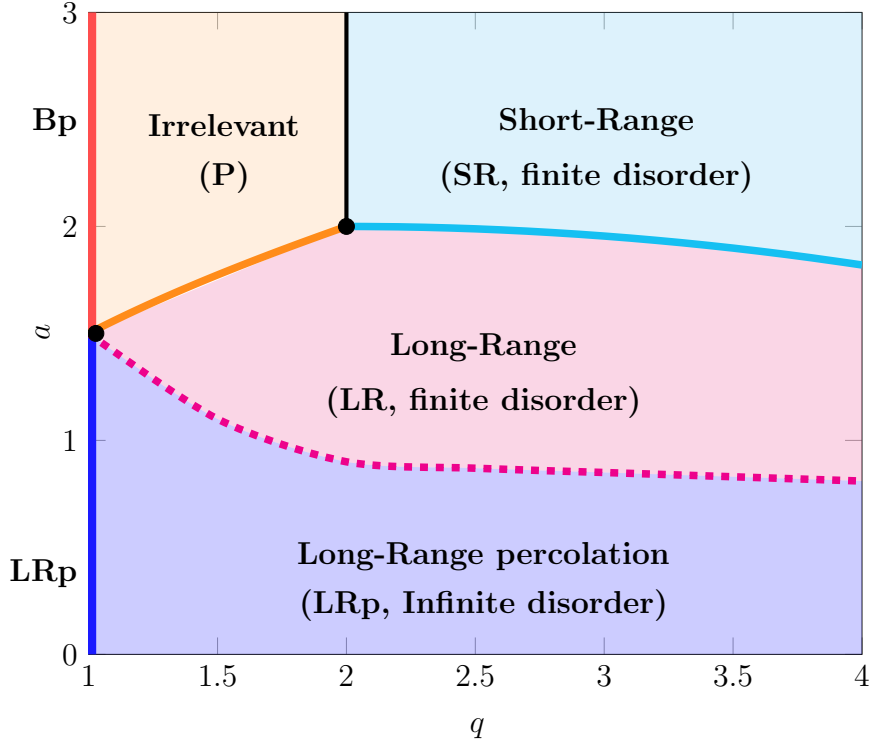
\begin{figure}
	\hspace*{1.5cm}
		\begin{tikzpicture}[scale=1.5]
\begin{axis}[title = {},
legend cell align=center,
xlabel={$q$},
x label style={at={(0.335,-0.0135)}},
ylabel={$a$},
y label style={at={(.04,0.335)}},
xmin=1,xmax=4,ymin=0,ymax=3]

\addplot[line width=0.2cm, name path=Bp, red!70, smooth,mark= none] coordinates {(1,1.5) (1,3)}; 

\addplot[line width=0.2cm, name path=LRp, blue!90, smooth,mark= none] coordinates {(1,0) (1,1.5)}; 

\path[name path=axis] (axis cs:0,0) -- (axis cs:4,0);

\path[name path=axis1] (axis cs:0,3) -- (axis cs:4,3);

\addplot[const plot, no marks, line width=0.55mm, black] coordinates {(2,2) (2,3)};

\addplot[line width=0.1cm, name path=SR, cyan!70, smooth,mark= none] 
coordinates{(2., 2.)(2.2, 1.9982)(2.4, 1.99279)(2.6, 1.98379)(2.8, 1.97118)(3., 1.95497)(3.2, 1.93515)(3.4, 1.91174)(3.6,1.88472)(3.8, 1.8541)(4.,1.81987)
};

\addplot[line width=0.1cm, name path=P, orange!90, smooth,mark= none] 
coordinates{
	(1, 1.5) (23/22, 1.52781)(12/11, 1.5549)(25/22, 1.58132)(13/11, 1.60712)(27/22, 1.63233)(14/11, 1.657)(29/22, 1.68117)(15/11, 1.70486)(31/22, 1.72811)(16/11, 1.75093)(3/2, 1.77337)(17/11, 1.79543)(35/22, 1.81715)(18/11, 1.83854)(37/22,
	1.85962)(19/11, 1.88042)(39/22, 1.90094)(20/11, 1.92121)(41/22, 1.94123)(21/11, 1.96103)(43/22, 1.98061)(2, 2)};
\filldraw (100,200) circle (2pt);

\addplot[line width=0.1mm, name path=LR, magenta!90, dashed, smooth,mark= none] 
coordinates{(1,1.5)(1.5,1.1)(2,.9)(2.5,0.87)(3,0.85)(3.5,0.83)(4,0.81)};

\addplot[magenta!50, opacity=0.4] fill between[of=SR and LR];

\addplot[blue!50, opacity=0.4] fill between[of=LR and axis];

\addplot[orange!30, opacity=0.4] fill between[of=P and axis1, soft clip={domain=0:2}];

\addplot[cyan!30, opacity=0.4] fill between[of=SR and axis1, soft clip={domain=1.999:4}];

\addplot[line width=0.1cm, name path=magenta1, magenta, dashed, smooth,mark= none] 
coordinates{(1,1.5)(1.5,1.1)(2,.9)(2.5,0.87)(3,0.85)(3.5,0.83)(4,0.81)};
\filldraw (2.8,150) circle (2pt);

\end{axis}

\draw (-0.3,4.5) node[above]{\textbf{Bp}};
\draw (-0.4,0.5) node[above]{\textbf{LRp}};
\draw (4.5,4.5) node[above]{\textbf{Short-Range}};
\draw (4.5,4) node[above]{\textbf{(SR, finite disorder)}};
\draw (1.2,4.5) node[above]{\textbf{Irrelevant}};
\draw (1.2,4) node[above]{\textbf{(P)}};
\draw(3.5,2.5) node[above]{\textbf{Long-Range}};
\draw(3.5,2) node[above]{\textbf{(LR, finite disorder)}};
\draw(3.5,0.65) node[above]{\textbf{Long-Range percolation}};
\draw(3.5,0.2) node[above]{\textbf{(LRp, Infinite disorder)}};

\end{tikzpicture}
\caption{Phase diagram of the disordered bi-dimensional $q$-Potts model for $(q,a)\in (0,4]\times (0,\infty)$.}
\label{ExtHarrfig}
\end{figure}
\end{center} 


\section{The long-range bond disordered $q$-Potts model }

 The model is defined by the partition function:
\begin{align}
\label{dpb}
\mathcal{Z}(\{J_{<ij>}\})= \sum_{\{s_i\}}\; \exp{\left(\sum_{<ij>} J_{<ij>} \delta_{s_i, s_j}\right)},
\end{align}
where the spin $s_i$, living on the lattice vertex $i$, takes $q$ possible states, $s_i = \left\{ 1,\cdots, q \right\}$. The $<ij>$ identifies the edge connecting the neighbouring sites $i$ and $j$ and the $\delta_{k,l}$ is the Kronecker delta. The set of couplings $\{J_{<ij>}\}$ is made of random variables drawn from a given distribution, the average over which will be indicated by the notation $\mathbb{E}[\cdots]$. 
 
We extract the critical exponent of the model~(\ref{dpb}) by measuring the properties of its Fortuyn-Kasteleyn (FK) clusters\cite{kasteleyn1969phase,fortuin1972random}. The FK clusters are the connected sets of bonds that enter in the geometric representation of the Potts partition function.
From Eq.~(\ref{dpb}), we get:
\begin{equation}
\begin{split}
\label{FKd}
\mathcal{Z}\left(\{J_{<ij>}\}\right) & =\; \sum_{\{s_i\}}\ \prod_{<ij>}\left[1 + (e^{J_{<ij>}}-1)\delta_{s_i, s_j}\right] \\
&=\sum_{\mathcal{G}}q^{\text{\#\;FK\;clusters($\mathcal{G}$)}}\left(\prod_{<ij>\in \mathcal{G}} e^{J_{<ij>}}-1\right) \\  
& = \sum_{\mathcal{G}} \prod_{<ij>\in \mathcal{G}}\left(1-e^{-J_{<ij>}}\right)\prod_{<ij>\notin \mathcal{G}}\left(e^{-J_{<ij>}}\right)\;q^{\text{\#\;FK\;clusters($\mathcal{G})$}},
\end{split}
\end{equation}
where the sum is over the set $\mathcal{G}$ of activated edges or bonds. Note that in the last equality  we discard a global factor $\prod \limits_{< ij >} e^{J_{< ij >}}$. A bond connects two equal spins with  probability $p_{<ij>}=1-e^{-J_{<ij>}}$. 

We will consider self-averaging observables, such as the thermal ($\nu$) and the order-parameter ($\beta$) critical exponents. More specifically, we will compute, for $q=1,2,3$ the ratio $\beta/\nu$ from the measure of the FK fractal dimension $d_f$ using:
\begin{equation}
\label{betadf}
d_{f}=2-\frac{\beta}{\nu},
\end{equation}
see Section~\ref{MCmeasures}.
The exponent $\nu$, instead, is extracted, for $q=3$, from the FK clusters wrapping probability, defined in Eq.~(\ref{def:wp}). 
The exponent $\nu$ determines the finite size scaling of the probability defined in Eq.~(\ref{eq:wrscaling}). We consider, in Section~\ref{sec:q2},  the Ising ($q=2$) spin-spin correlation function, which is a non self-averaging observable. We will measure the multifractal behaviour of this quantity, see Eq.~(\ref{def:ss}), and compare with recent predictions \cite{dudka2016critical}. 
\subsection{Implementation of disorder}
\label{numsetup}
In our simulations, we implement a quenched bi-modal disorder where the couplings can take randomly two values, $J_{<ij>}=J_1$ or $J_{<ij>}=J_2$, with equal probability.
The values of $J_1$ and $J_2$ are not taken independently but they are chosen to satisfy the following self-dual condition:
\begin{align}
\label{dual}
(e^{J_1}-1) (e^{J_2} -1 ) = q.
\end{align}
The self-dual line Eq.~(\ref{dual}) corresponds to a line of critical points separating a paramagnetic phase and a ferromagnetic phase \cite{kinzel1981critical}.  

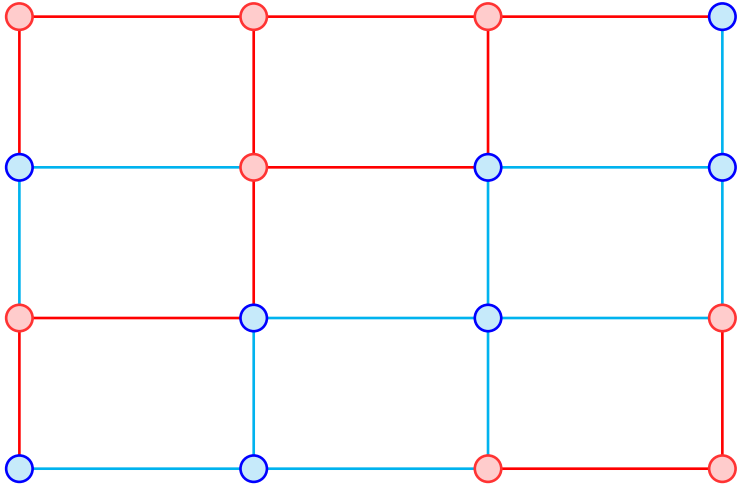
\begin{figure}[!ht]
 	\centering
 	\begin{tikzpicture}[scale = 2.5,  every node/.style={scale=2.5}]
 	\begin{axis}[
 	hide x axis,
 	hide y axis,
 	scaled x ticks=manual:{}{\pgfmathparse{#1}},
 	scaled y ticks=manual:{}{\pgfmathparse{#1}},
 	tick align=outside,
 	tick pos=both,
 	x grid style={white!69.0196078431373!black},
 	xmin=-1.0, xmax=4.5,
 	xtick style={color=black},
 	xticklabels={},
 	y grid style={white!69.0196078431373!black},
 	ymin=-4, ymax=3.1,
 	ytick style={color=black},
 	yticklabels={}
 	]
 	
 	\path [draw=cyan!90]
 	(axis cs:1,1)
 	--(axis cs:1,0);
 	
 	\path [draw=cyan!90]
 	(axis cs:1,1)
 	--(axis cs:2,1);
 	
 	\path [draw=cyan!90]
 	(axis cs:2,1)
 	--(axis cs:2,0);
 	
 	\path [draw=cyan!90]
 	(axis cs:3,2)
 	--(axis cs:3,1);
 	
 	\path [draw=cyan!90]
 	(axis cs:1,0)
 	--(axis cs:2,0);
 	
 	\path [draw=cyan!90]
 	(axis cs:3,3)
 	--(axis cs:3,2);
 	
 	\path [draw=cyan!90]
 	(axis cs:0,0)
 	--(axis cs:1,0);
 	
 	\path [draw=cyan!90]
 	(axis cs:0,1)
 	--(axis cs:0,2);
 	
 	\path [draw=cyan!90]
 	(axis cs:0,2)
 	--(axis cs:1,2);
 	
 	\path [draw=cyan!90]
 	(axis cs:2,2)
 	--(axis cs:3,2);
 	\path [draw=cyan!90]
 	(axis cs:2,2)
 	--(axis cs:2,1);
 	\path [draw=cyan!90]
 	(axis cs:2,1)
 	--(axis cs:3,1);
 	\path [draw=red]
 	(axis cs:0,0)
 	--(axis cs:0,1);
 	
 	\path [draw=red]
 	(axis cs:0,1)
 	--(axis cs:1,1);
 	
 	\path [draw=red]
 	(axis cs:1,1)
 	--(axis cs:1,2);
 	
 	\path [draw=red]
 	(axis cs:1,2)
 	--(axis cs:2,2);
 	
 	\path [draw=red]
 	(axis cs:1,2)
 	--(axis cs:1,3);
 	
 	\path [draw=red]
 	(axis cs:1,3)
 	--(axis cs:0,3);
 	
 	\path [draw=red]
 	(axis cs:0,3)
 	--(axis cs:0,2);
 	
 	\path [draw=red]
 	(axis cs:1,3)
 	--(axis cs:2,3);
 	
 	\path [draw=red]
 	(axis cs:2,3)
 	--(axis cs:3,3);
 	
 	\path [draw=red]
 	(axis cs:2,3)
 	--(axis cs:2,2);
 	
 	\path [draw=red]
 	(axis cs:2,0)
 	--(axis cs:3,0);
 	
 	\path [draw=red]
 	(axis cs:3,0)
 	--(axis cs:3,1);
 	
 	\addplot [only marks, mark=*, draw=blue, fill=cyan!20, colormap/viridis]
 	table{%
 		x                      y
 		0 0
 		0 2
 		1 0
 		1 1
 		2 1
 		2 2
 		3 2
 		3 3
 		
 	};
 	\addplot [only marks, mark=*, draw=red!80, fill=red!20, colormap/viridis]
 	table{%
 		x                      y
 		
 		0 1
 		0 3
 		1 2
 		1 3
 		2 0
 		2 3
 		3 0
 		3 1
 	};
 	
 	\end{axis};
 	\end{tikzpicture}
 	\vspace*{-80mm}
 	\caption{A particular configuration $\{\sigma_i\}$ is shown. Blue (red) filled circles are the sites corresponding to $\sigma=1$ ($\sigma=-1$). The corresponding set of couplings $J_{<ij>}$, associated to the edges, follow from the Eq.~(\ref{bimodal}). In the Figure, the $J_{<ij>}=J_1$ ($J_{<ij>}=J_2$) couplings are associated to the blue (red) colored edges.}
 	\label{sigmaconf}
 \end{figure}

In order to simulate this disorder, we associate to each vertex of the square lattice a random variable  $\sigma_i=\{-1,1\}$ and we set:  
\begin{align}
\label{bimodal}
J_{<ij>^{(R)}} =J_{<ij>^{(B)}}= \frac{J_1+J_2}{2} + \sigma_i \frac{J_1 - J_2}{2},
\end{align}
 where, the $J_{<ij>^{(R)}}$ and $J_{<ij>^{(B)}}$ are the couplings associated respectively to the edge on the right and to the edge on the bottom of a given site $i$. The Fig.~(\ref{sigmaconf}) illustrates the connection between the $\sigma_i$ and the $J_{<ij>}$ configurations as given by the Eq.~(\ref{bimodal}).
 
In principle there are other possibles choices: for instance one could introduce, instead, an edge random variable $\sigma_{<ij>}$ and impose $J_{<ij>}= (J_1+J_2)/2 + \sigma_{<ij>} (J_1 - J_2)/2$. All these different set-ups are expected to be equivalent as far universal properties are concerned. We have chosen simply the one that was more convenient for our numerical simulations. 
 
The disorder distribution is then fixed by the $\{\sigma_i\}$ one. Here we consider the distributions that are characterized by a
vanishing first moment:
\begin{equation}
\label{1cum}
\mathbb{E}\left[\sigma_i\right]=0,
\end{equation}
which forces the couplings to take values $J_1$ or $J_2$ with the same probability. The couplings then fluctuate around the value $(J_1+J_2)/2$:
\begin{equation}
J_{<ij>}=\mathbb{E}[J_{<ij>}]+\delta J_{<ij>}, \quad \mathbb{E}[J_{<ij>}]=\frac{J_1+J_2}{2}.
\end{equation}
We can then consider a $\{\sigma_i\}$ distribution whose second cumulant has a power-law decaying:
\begin{align}
\label{defa}
&\mathbb{E}\left[\sigma_i\;\sigma_k\right]\sim  |i-k|^{-a}\quad \text{for}\; |i-k|\gg1.
\end{align}
This, of course, correlates the fluctuations of the couplings at distant edges:
\begin{align}
\mathbb{E}\left[\delta J_{<ij>^{(X)}}\;\delta J_{<kl>^{(Y)}}\right]&=\frac{(J_1-J_2)^2}{4}\mathbb{E}\left[\sigma_i\;\sigma_k\right]\nonumber \\
&\sim  |i-k|^{\displaystyle -a} \quad \text{for}\; |i-k|\gg1,
\end{align}
where $X,Y=R,B$. 

Once we have chosen the $\{\sigma_i\}$ distribution and satisfied the relation Eq.~(\ref{dual}), the phase diagram of the $q$-Potts model Eq.~(\ref{dpb}) depends only on the ratio $r$, 
\begin{equation}
\label{defr}
r=\frac{J_1}{J_2}\in [1,\infty],
\end{equation} 
that parametrizes the disorder strength. We assumed, without loss of generality, that $J_2\leq J_1$. In our simulations we will analyse the behaviour of the system by varying $r$. The stability of the RG fixed points are therefore probed  with respect to this direction while the relation Eq.~(\ref{dual}) makes the system stay at the critical temperature. 

\subsection{Pure Potts model (r=1)}
\label{sec:r1}


The pure case corresponds to $r=1$ where the couplings take all the same value, see Eq.~(\ref{defr}). On the square lattice, the corresponding critical point P is located at $ J_1=J_2=J_c= \log{(1+\sqrt{q})}$.

In the continuum limit, the P point is described by the Potts conformal field theory whose energy spectrum is known since longtime \cite{fsz87}. In these last years a bootstrap approach \cite{picco2013connectivities,ei15,prs16,prs19,ribault22} has been proposed for studying the cluster connectivity properties, which finally was used in \cite{js18,NivesvivatRibault,jrs22}. This provides the exact Potts correlation functions and a complete characterization of the symmetry representation of its states. 

The thermal and order-parameter critical exponent $\nu^{P}$ and $\beta^{P}$ of the pure Potts model have been determined for any value of $q\in [0,4]$, see for instance \cite{henkel2012conformal2} and references therein. In Table~\ref{Ptable}, we give the values of $\nu ^{P}$ and of $\beta^P/\nu^{P}$ for $q=1,2,3$:

\begin{table}
\centering
\renewcommand{\arraystretch}{1.5}
\footnotesize

\begin{tabular}{|p{.5cm}|P{.5cm}|P{.9cm}|}
\hline
$q$ \hspace{0.5cm}&$\nu^{P}$\hspace{0.5cm} & $\beta^{P}/\nu^P$\hspace{0.5cm} \\ & &\\
\hline 
$1$ & $4/3$ & $5/48$\\ & &\\
\hline
2&$1$& $1/8$\\ & &\\
\hline
3 & $5/6$ & $2/15$\\& & \\
\hline

\end{tabular} 
\caption{Pure critical exponents for different $q$ values. }
\label{Ptable}
\end{table}

\subsection{Short-range disordered Potts ($a\geq 2$)}
\label{SRD}
When $a\geq 2$ the disorder is effectively uncorrelated. In Fourier space, the contribution of the power-law tail $|x|^{-a}$  is of order $O(k^{a-2})$ and, in the long wavelength limit $k\to 0$, is finite for $a\geq 2$. At large distances, the properties of a long-range distribution are the same as the one of a delta function.   
   
The phase diagram of the uncorrelated disorder is quite well understood.
For $q<2$  the P  point is the only stable RG point: the weak disorder does not modify the universality of the pure model. For $q>2$ instead, the system is driven to a stable SR point. For $q=2$ the disorder is marginal (the P and SR points coincide) and generate universal logarithmic corrections to the P universality class \cite{Dotsenko_1982}. We report in Table~\ref{SRtable} the critical exponents for $q=3$, computed perturbatively in \cite{ludwig1990infinite,dotsenko1995renormalisation}:
\begin{table}[!ht]
\centering
\renewcommand{\arraystretch}{1.5}
\footnotesize

\begin{tabular}{|p{.4cm}|P{1.1cm}|P{1.9cm}|}

\hline
$q$ \hspace{0.5cm}& $\nu^{SR}$\hspace{0.5cm} & $\beta^{SR}/\nu^{SR}$\hspace{0.5cm} \\ & &\\
\hline 
3 & $\sim 1.02$ & $\sim 0.134655 $\\& & \\
\hline 
\end{tabular} 
\caption{Short-Range critical exponents for $q=3$.}
\label{SRtable}
\end{table}

In the case of uncorrelated disorder, one can show by simple power counting that only the first two cumulants of the distributions determine the long-distance behaviour of the system \cite{dotsenko1995renormalisation}.

\subsection{Long-Range disordered Potts ($a<2$)}
\label{CD}
For $a<2$ one has to take into account the power-law decay of the disorder distribution.
As mentioned in the Introduction, according to the extended Harris criterion \cite{WeinribHalperin,Weinrib}, for $(q<2, a<2/\nu^{P})$ and for $(q>2, a<2/\nu^{SR})$, the long-range part of the distribution is relevant and dominates the short-range one.
In \cite{WeinribHalperin,Weinrib}, it was also observed that, when it exists, the LR point has to be stable  with respect to an additional  sub-dominant term to $\mathbb{E}[\delta J(x)\delta J(y)]$, $\mathbb{E}[\delta J(x)\delta J(y)]\to \mathbb{E}[\delta J(x)\delta J(y)]+ w_0 |x-y|^{-b}$, with $b>a$. Reiterating the extended Harris criterion, one expects that when $b>a$ then $b> 2/\nu^{LR}$, i.e. the perturbation is irrelevant. Otherwise, when $b<a$, the additional b-term becomes dominant and therefore $b<2/\nu^{LR}$. This in turn brings to the strong conjecture that: 
\begin{equation}
\label{nulr}
\nu^{\text{LR}}= \frac{2}{a}, \quad \text{for any}\;q.
\end{equation} 
The above relation has been proven by perturbative RG computation in different long-range disordered models  \cite{WeinribHalperin,Weinrib,Rajabpour_2007,dudka2016critical} at one or two-loops order. It is now also quite established that the Eq.~(\ref{nulr}) should be valid at all perturbation orders \cite{Honkonen_1989,Korzhenevskii_1995}. In Section~\ref{sec:nu}, we will test the relation Eq.~(\ref{nulr}) for $q=3$.

\subsection{The infinite disorder point ($r=\infty$): the $q$-colored critical pure percolation}
\label{sec:rinfty}
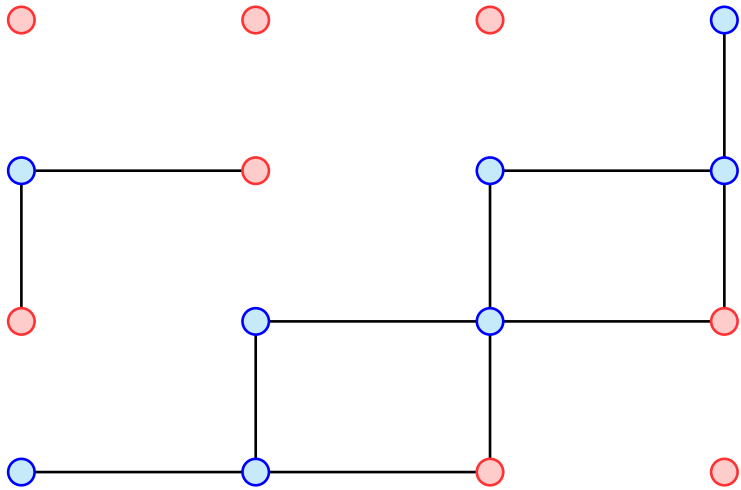
\begin{figure}[!ht]
	\centering
	\begin{tikzpicture}[scale = 2.5,  every node/.style={scale=2.5}]
	\begin{axis}[
	hide x axis,
	hide y axis,
	scaled x ticks=manual:{}{\pgfmathparse{#1}},
	scaled y ticks=manual:{}{\pgfmathparse{#1}},
	tick align=outside,
	tick pos=both,
	x grid style={white!69.0196078431373!black},
	xmin=-1.0, xmax=4.5,
	xtick style={color=black},
	xticklabels={},
	y grid style={white!69.0196078431373!black},
	ymin=-4, ymax=3.1,
	ytick style={color=black},
	yticklabels={}
	]

	\path [draw=black]
	(axis cs:1,1)
	--(axis cs:1,0);
	
	\path [draw=black]
	(axis cs:1,1)
	--(axis cs:2,1);
	
	\path [draw=black]
	(axis cs:2,1)
	--(axis cs:2,0);
	
	\path [draw=black]
	(axis cs:3,2)
	--(axis cs:3,1);
	
	\path [draw=black]
	(axis cs:1,0)
	--(axis cs:2,0);
	
	\path [draw=black]
	(axis cs:3,3)
	--(axis cs:3,2);
	
	\path [draw=black]
	(axis cs:0,0)
	--(axis cs:1,0);
	
	\path [draw=black]
	(axis cs:0,1)
	--(axis cs:0,2);
	
	\path [draw=black]
	(axis cs:0,2)
	--(axis cs:1,2);
	
	\path [draw=black]
	(axis cs:2,2)
	--(axis cs:3,2);
	\path [draw=black]
	(axis cs:2,2)
	--(axis cs:2,1);
	\path [draw=black]
	(axis cs:2,1)
	--(axis cs:3,1);
	%
	%
	%
%
%
%
%
%
%
%
%
%
%
%
%
	
	\addplot [only marks, mark=*, draw=blue, fill=cyan!20, colormap/viridis]
	table{%
		x                      y
		0 0
		0 2
		1 0
		1 1
		2 1
		2 2
		3 2
		3 3
		
	};
	\addplot [only marks, mark=*, draw=red!80, fill=red!20, colormap/viridis]
	table{%
		x                      y
		
		0 1
		0 3
		1 2
		1 3
		2 0
		2 3
		3 0
		3 1
	};
	
	\end{axis};
	\end{tikzpicture}
	\vspace*{-80mm}
\caption{Given the configuration of $\sigma_i$ and therefore of $J_{<ij>}$ shown in Fig.~(\ref{sigmaconf}), we  show the only random cluster graph $\mathcal{G}$ in (\ref{FKd}) that does not vanish in the limit $r\to \infty$.  Notice that the $\sigma_i=1$ cluster, formed of blue spins, are different from the ones of the FK bonds (black lines)}
\label{Fig:rinfty}
\end{figure}

The $r \rightarrow \infty$ point is the only point,  besides the pure one ($r=1$), where exact results can be given for any $q$.
From the Eq.~(\ref{dual}), in the limit $r \rightarrow \infty$ one has  $J_1 \rightarrow \infty$ and $J_2 \rightarrow 0$ with 
$r J_2\sim -\log(J_2)\to \infty$. In particular:
\begin{equation}
\mathbb{E}\left[\delta J_{<ij>}\delta J_{<kl>} \right]= \frac{(r-1)^2 J_2^2}{4} \mathbb{E}\left[\sigma_i\sigma_k\right]\sim (r J_2)^2  \mathbb{E}\left[\sigma_i\sigma_k\right] \; .
\end{equation}
The $r=\infty$ limit corresponds then to the maximal amount of disorder we can consider. In this limit, the Eq.~(\ref{FKd}) greatly simplifies. 
Indeed, with probability one, there is a bond on an edge with coupling  $J_1$ ($1-e^{-J_1}=1$) and probability zero of having a bond on a edge with $J_2$ ($1-e^{-J_2}=0$). So, once a configuration of  $\sigma_{i}$ (and consequentely of the couplings $J_{<ij>}$) is drawn, only the graph where the edges $<ij>$ associated to $J_1$ are activated, contributes to the partion sum, see Fig.~(\ref{Fig:rinfty}). The $J_2$ edges are not activated.

The  disorder average free-energy, for instance, can be written as:
\begin{equation}
\label{rinfty}
\mathbb{E}\left[\log{\mathcal{Z}(\{J_{<ij>}\}}\right]=\log{q} \times \mathbb{E}\left[\text{\# $\left(J_1 \text{-clusters} \right)$}\right]\sim \;\log{q} \times \mathbb{E}\left[\text{\# $\left(\sigma=+1 \text{-clusters} \right)$}\right]
\end{equation}
where $\sim$ means that the scaling behaviour is the same.  The disordered Potts model at $r = \infty$ is then strictly related to the site percolation model based on the $\sigma$ clusters. In particular, the probability distribution of the $r=\infty$ FK clusters is, besides a trivial dependence on $q$, the same as the one of the $\sigma$ clusters. In our construction Eq.~(\ref{bimodal}), the $<ij^{(R)}>$ and $<ij^{(B)}>$ are always activated or not activated at the same time. At the lattice level, the $r=\infty$ FK clusters are therefore different from the $\sigma$ clusters, as illustrated in Fig.~(\ref{Fig:rinfty}). However their universal behaviour, such as their fractal dimension, as well as the correlation lenght exponent $\nu$, remains the same, as discussed in the following Section~\ref{MCrinfty}. A more quantitative argument can be provided for  $a=1/4$, where the $\sigma$ are Ising spins at the critical point. In fact, one can observe that the two type of clusters differ only on the boundaries. For an Ising cluster of linear size $R$, the number of spins on the boundary scales as $\simeq R^{d^b_f}$ with $d^b_f = 11/8=1.375$\footnote{Note that the Ising spin boundaries are described by the CLE$_{3}$ loops and $d^{b}_f=11/8$ is the fractal dimension of the CLE$_{3}$ loops \cite{sheffieldwerner}}, which is much smaller than the number of spins in the cluster that scales as $R^{d_f}$ with $d_f = 187/96 \simeq 1.948$. Then, the FK clusters for $a=0.25$ are expected 
to have the same fractal dimension $d_f = 2 - 5/96$ as the Ising spin clusters.

When $a>2$, the system is short-range and it is described by the Bernoulli critical point Bp\cite{cardy1997critical}. For $a<2$, one applies the extended Harris criterion using $\nu^{Bp}=4/3$: for $a>2/\nu^{Bp}=3/2$ the system remains in the Bp point while $a<3/2$, there is the new LRp fixed point already mentioned in the Introduction. 
In Section~\ref{MCrinfty} we tested this behaviour and we computed the exponent $\beta^{LRp}/\nu^{LRp}$. 

We insist on the fact that the LRp exists at $r = \infty$ for any $q$ and its critical exponents depend on $a$ but do not depend on $q$. As we will discuss in the following, see Fig.s~(\ref{Figq2}, \ref{Figq3}), the LRp point is for $q>1$ distinct from the LR point, which is the finite disorder fixed point. The LRp point exhanges stability with the LR one when $a$ is greater than a particular value $a^*(q)$, $a>a^*(q)$.

\section{Phase diagrams from Monte-Carlo measures of $q=1,2,3$-Potts}
\label{MCmeasures}

We present here the numerical results according to which we established the phase diagram of Fig.~(\ref{ExtHarrfig}). Simulations were done on square lattices with periodic boundary conditions. If not stated otherwise, we average over one million samples of disorder. For each value of $q$ and $a$, we determined first the average autocorrelation time 
$\tau_{a,q}(L)$. Next the thermal average is done, for each sample of disorder, by runing $100 \times \tau_{a,q}(L)$ updates after the same number of updates for thermalisation. 
If not stated otherwise, the results presented in the following are obtained by using the disorder distribution described in Section~\ref{nising}.

For each value of $n=\{1,\cdots, 8\}$, corresponding to values $a$ given in Eq.~(\ref{avsn}), we generate the coupling configuration $J_{<ij>}$, see Eq.~(\ref{bimodal}).
We then construct the FK clusters by connecting equal spins with the probability $p_{<ij>} = 1 - e^{-J_{<ij>}}$, see Eq.~(\ref{FKd}). Given a lattice of linear size $L$, we compute, for each coupling configuration, the (thermal averaged) number of spins  $M (\{J_{<ij>}\},L)$ of the largest FK cluster. We average on the coupling distribution, 
\begin{equation}
M(L)= \mathbb{E}\left[M (L,\{J_{<ij>}\})\right],
\end{equation}
to extract the fractal dimension $d_f$ by the scaling
\begin{align}
\label{scalingdf}
M (L) \simeq L^{d_f}.
\end{align}
More specifically, recalling the relation Eq.~(\ref{betadf}), we use Eq.~(\ref{scalingdf}) to compute an effective magnetic dimension:
\begin{equation}
\label{b_nu_numerics}
\frac{\beta}{\nu} (L)= -\log \left[\displaystyle \frac{m(L)}{m(L/2)}\right]/\log\left[2\right],
\end{equation}
where $m(L)=M(L)/L^2$ is the average magnetization. The above data, computed for different $q$, $a$ and $r$ are the ones shown in the plots $\beta/\nu$ vs. $L$ below.   
The ratio $\beta/\nu$ is then evaluated in the scaling limit:
\begin{equation}
\frac{\beta}{\nu}= \lim_{L\to \infty}\;\frac{\beta}{\nu} (L)\; .
\end{equation}

%
\subsection{The infinite disorder point ($r=\infty$): LRp and Bp fixed points}
\label{MCrinfty}
We collect here results for the $q$-colored percolation model, found at the infinite disorder point $r=\infty$, see Section~\ref{sec:rinfty}. We recall that the critical exponents at the infinite disorder point do not depend on the value of $q$.
\begin{figure}[!ht]
\begin{center}
\includegraphics[width=12cm,height=10cm]{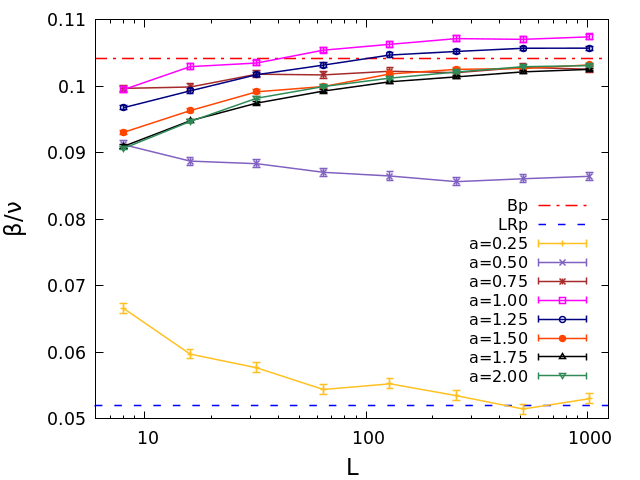}
\end{center}
\caption{The $r=\infty$ long-range disorder: $\beta/\nu$ vs. $L$ for the values of $a$ shown in the caption.}
\label{Fig2}
\end{figure}

In Fig.~(\ref{Fig2}), we show the effective exponent $\beta/\nu(L)$ 
as a function  of $L$ for $a=\{0.25, \cdots, 2.00\}$. 
The corresponding scaling limit results $\beta^{LRp}/\nu^{LRp}$, which are obtained by a best fit of all our data, including a sub-leading correction, are reported in Table~\ref{LRptable}. 

\begin{table} [!ht]
	\centering
	\renewcommand{\arraystretch}{1.5}
	\footnotesize
	\centering
	
	\begin{tabular}{|P{2.7cm}|P{1.7cm}|}
	\hline
	$a$ &  $\beta^{LRp}/\nu^{LRp}$  \\
	\hline 
	$0.25$ & $0.0522 (4)$  \\
	\hline 
	$0.5$ & $0.086 (2) $  \\
	\hline
	$0.75$& $0.103 (2)$  \\
	\hline
	$1,1.25,1.5,1.75,2$ & $ \sim 0.105$  \\	
	\hline
\end{tabular}
\caption{Best fit of the critical exponent $\beta^{LRp} / \nu^{LRp}$ for different values of $a$. Notice that for $a=\{1.5,1.75,2\}$ the system is described by the Bp point where $\beta^{Bp}\ / \nu^{Bp} = 5/48 \sim 0.105.$ However, for $a\geq 1$, the numerical simulations cannot distinguish between the LRp point and the Bp point.}
\label{LRptable}
\end{table}

As discussed in Section~\ref{sec:rinfty}, for $a=0.25$, the $\beta^{LRp}/\nu^{LRp}$ must correspond to the fractal dimension of the Ising clusters. The latter has been considered in \cite{Coniglio1980,vanderzande92} and argued to be:
\begin{equation}
\label{coniglio}
\frac{\beta^{LRp}}{\nu^{LRp}}=2-\frac{187}{96}=\frac{5}{96}\sim 0.0521.
\end{equation}
From the Table~\ref{LRptable}, one can see that the agreement to the value $5/96$, shown as a dashed line in Fig.~(\ref{Fig2}), is very good.

We remind that according to the Harris criterion, one expects that, for $a<3/2$, $\beta^{LRp}/\nu^{LRp}<\beta^{Bp}/\nu^{Bp}$. As a matter of fact, the difference between  $\beta^{LRp}/\nu^{LRp}$ and $\beta^{Bp}/\nu^{Bp}$ becomes very small for increasing $a$ and the numerical simulations cannot effectively distinguish between these two fixed points. In fact, for $a=1$ and $a=1.25$, it seems to be even slightly larger than $5/48$. A similar results was also observed in \cite{Janke_2017}. But, since the finite size correction 
are important, it is difficult to really predict the asymptotic value. 


We consider in the following Potts observables at finite value of disorder $r$, where the scaling behaviour is in general expected to depend on $q$.

\subsection{The $q=1$ phase diagram}
\label{sec:q1}
We consider here the phase diagram of the $q=1$ Potts model. 
For $a>2$ the disorder is effectively short-range and irrelevant, see Fig.~(\ref{ExtHarrfig}). The fact that a short-range disorder is irrelevant can also be explained  by a simple lattice argument.  On the critical line Eq.~(\ref{dual}), for $r=1$,  $J_1=J_2=J_c=\log{(1+\sqrt{2})} = \log 2$, thus we are exactly at the Bp critical point, $p=1-e^{-J_c}=p_c=1/2$.  
By adding disorder, meaning by setting $r>1$, the dual line Eq.~(\ref{dual}) fixes, at $q=1$, the relation $1-e^{-J_1} - e^{-J_2} = 0$. 
Recalling that each bond $J_1$ and $J_2$ occurs with equal probability, the density of activated bond remains $p=\frac12 (1-e^{-J_1}) + \frac12 (1 - e^{-J_2}) = 1/2$. So, we can conclude that, for any value of $r$, we stay in the Bp point. 
For $3/2=2/\nu^{Bp}<a<2$, the long-range disorder remains irrelevant, and the system is still described by the Bp point. 

The effective exponent $\beta/\nu (L)$ is measured for various values of disorder strength  $r$. For $a<3/2$ our numerical measures shows clearly that the systems flows from the Bp point to the LRp point, located at $r= \infty$. The LRp point and its scaling exponents have been discussed in the Section~\ref{sec:rinfty} and in Section.~\ref{MCrinfty}.

\begin{figure}[!ht]
\begin{center}
\includegraphics[width=16cm,height=7cm]{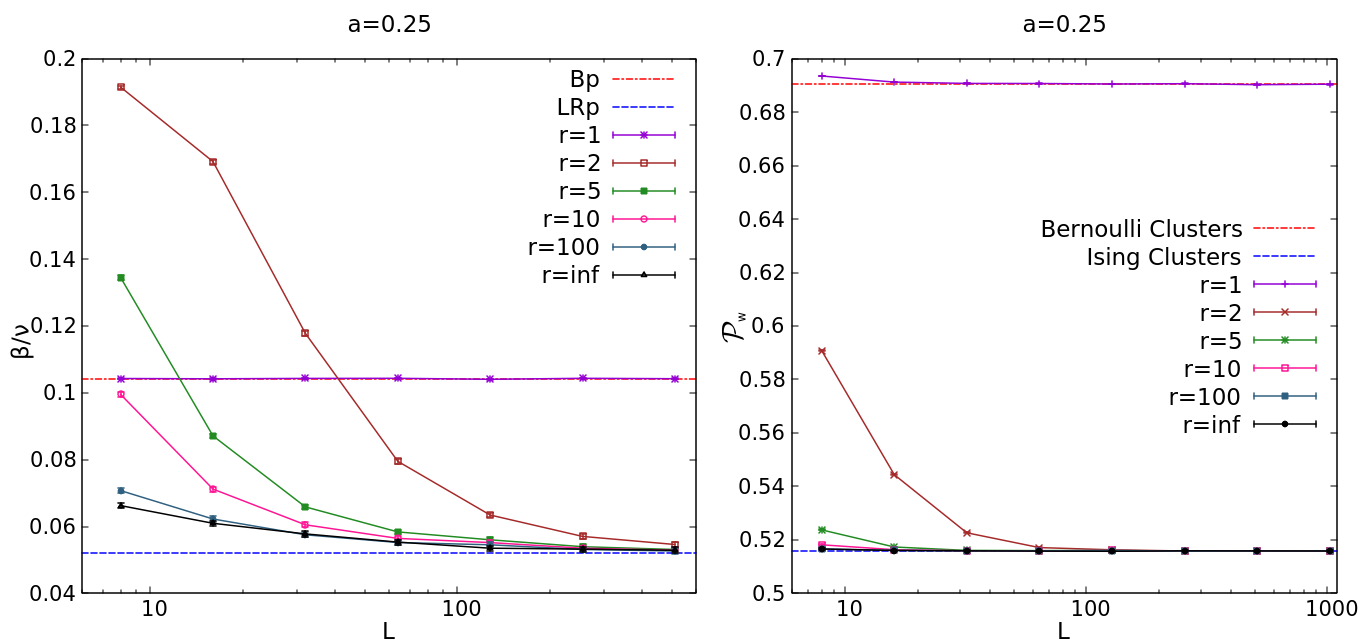}
\end{center}
\caption{$q=1$ Potts model with $a=0.25$. The left panel shows $\beta/\nu$ vs. $L$ while the right panel shows the wrapping probability ${\cal P}_w(L)$ vs. $L$.} 
\label{Fig1_a0.25}
\end{figure}

We present first the model with $a=1/4$. In the left part of Fig.~(\ref{Fig1_a0.25}), $\beta/\nu (L)$ is shown for $r=1, 2, 5, 10, 100$ and $\infty$ as a function of the size $L$. We observe that this exponent, for all values of $r>1$, goes towards the value of $r=\infty$, with a crossover function of the disorder. The corresponding expected value $5/96$, see \cite{Coniglio1980}, is shown as a blue dashed line. The $r=1$ corresponds to Bernoulli percolation whose scaling limit value $\beta^{Bp}/\nu^{Bp}=5/48$ is shown as a red dashed-dotted line. Thus, for any amount of disorder, the large distance behaviour is controlled by an infinite disorder fixed point. 
In the right part of Fig.~(\ref{Fig1_a0.25}), we show ${\cal P}_w(L)$, see Eq.~(\ref{def:wp}). For $r=1$, it converges towards 
the exact value for the Bernoulli percolation \cite{Pinson1994}:
\begin{equation}
\label{pinson}
\text{Bernoulli clusters:}\quad \lim_{L\to\infty} {\cal}P_w(L) =0.69046
\end{equation}
which is shown as a red dashed-dotted line. For any value $r>1$, it goes to the same value as for LRp point with a crossover function of the disorder.  
In addition to the fractal dimension, we can verify that also the wrapping probability supports the fact that, as stated above, the FK clusters behave, at the LRp point, as the Ising spin ones. For spin clusters this probability can be computed from the results in \cite{Blanchard} and is:
\begin{equation}
\label{blanchard}
\text{Ising clusters:}\quad \lim_{L\to\infty} {\cal}P_w(L) =0.515884 \; .
\end{equation}
This is shown as a blue dashed line in the right part of Fig.~(\ref{Fig1_a0.25}). The agreement is perfect.

%

\begin{figure}[!ht]
	\begin{center}
	\includegraphics[width=16cm,height=7cm]{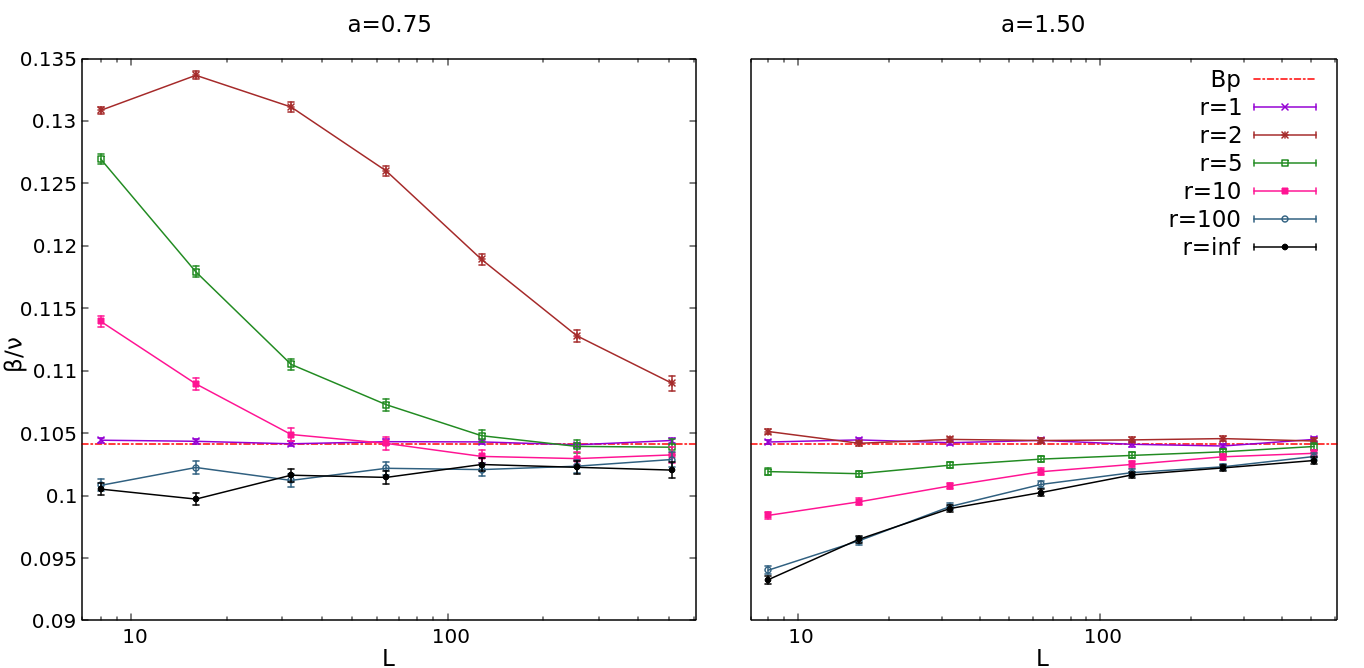}
	\end{center}
	\caption{$\beta/\nu$ vs. $L$ for the $q=1$ Potts model with long-range disorder with $a=0.75$ in the left panel and with $a=1.50$ in the right panel.}
	\label{Fig1_a0.75}
\end{figure}

Coming back to the $\beta/\nu$ exponent, similar behaviours are observed for the others values of $a$. In Fig.~(\ref{Fig1_a0.75}), we show $\beta/\nu(L)$ for $a=0.75$ and $a=1.5$. Again, for all values of $r$, the large size limit is the one reported in Table~\ref{LRptable}.

%

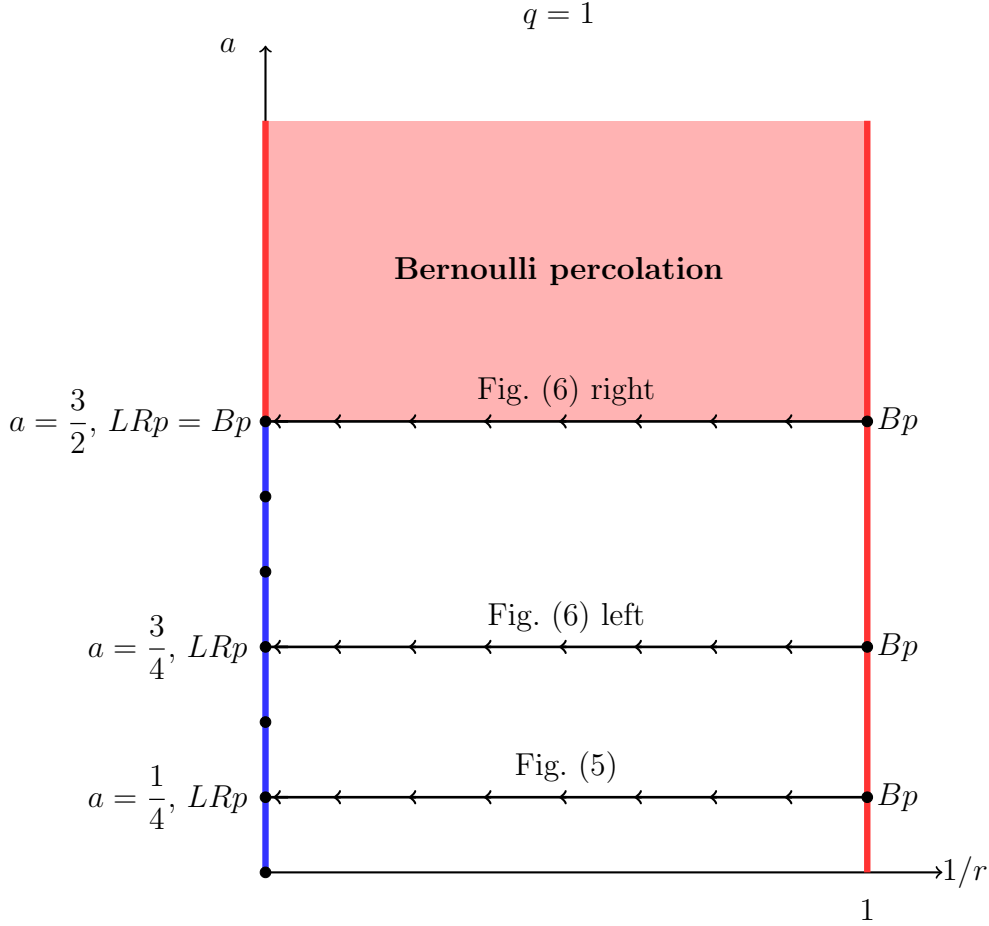
\begin{figure}[!ht]
 	\hspace*{1.2cm}
 	\scalebox{1.}{%
 	\begin{tikzpicture}[scale=1.00]
 	\draw[ thick]  [->](0.0,-2.5) -- (9.0,-2.5);
 	\draw[ thick]  [->](0.0,-2.5) -- (0.0,8.5);
 	
 	\node at (3.9,8.9) {$q=1$};
 	\node at (-0.5,8.5) {$a$};
 	\node at (9.3,-2.5) {$1/r$};
 	\node at (8.,-3.) {$1$};

 	\fill[color=red!30]
 	(0.02,3.5) -- (7.96,3.5)
 	-- (7.96,7.5) -- (0.02,7.5) -- cycle;
 	\node at (3.9,5.5) {\textbf{Bernoulli percolation}};
 	
 	\draw[ line width=0.85mm, , red!80]  (8.0,-2.5) -- (8.0,7.5);
 	\draw[ line width=0.85mm, , red!80]  (0.0,3.5) -- (0.0,7.5);
 	
%
%
%
%
 	
 	\node at (-1.8,3.5) {$\displaystyle a=\frac{3}{2} \text{, } LRp=Bp$};
 	
 	\draw[ line width=0.35mm, , black]  (0.0,3.5) -- (0.3,3.5);
 	\draw[<-][ line width=0.35mm, , black]  (0.1,3.5) -- (1.0,3.5);
 	\draw[<-][ line width=0.35mm, , black]  (0.9,3.5) -- (2.0,3.5);
 	\draw[<-][ line width=0.35mm, , black]  (1.9,3.5) -- (3.0,3.5);
 	\draw[<-][ line width=0.35mm, , black]  (2.9,3.5) -- (4.0,3.5);
 	\draw[<-][ line width=0.35mm, , black]  (3.9,3.5) -- (5.0,3.5);
 	\draw[<-][ line width=0.35mm, , black]  (4.9,3.5) -- (6.0,3.5);
 	\draw[<-][ line width=0.35mm, , black]  (5.9,3.5) -- (7.0,3.5);
 	\draw[<-][ line width=0.35mm, , black]  (6.9,3.5) -- (8.0,3.5);	
 	\draw[ line width=0.85mm, , blue!80]  (0.0,3.5) -- (0.0,-2.5);
 	\filldraw (0.,-2.5) circle (2pt);
 	\filldraw (0.,3.5) circle (2pt);
 	\node at (8.4,3.5) {$Bp$};
 	\node at (4.0,3.9) {Fig.~(\ref{Fig1_a0.75}) right};
 	\filldraw (8.,3.5) circle (2pt);
 	%

 	\filldraw (0,2.5) circle (2pt);

 	\draw[ line width=0.35mm, , black]  (0.0,0.5) -- (0.3,0.5);
 	\draw[<-][ line width=0.35mm, , black]  (0.1,0.5) -- (1.0,0.5);
 	\draw[<-][ line width=0.35mm, , black]  (0.9,0.5) -- (2.0,0.5);
 	\draw[<-][ line width=0.35mm, , black]  (1.9,0.5) -- (3.0,0.5);
 	\draw[<-][ line width=0.35mm, , black]  (2.9,0.5) -- (4.0,0.5);
 	\draw[<-][ line width=0.35mm, , black]  (3.9,0.5) -- (5.0,0.5);
 	\draw[<-][ line width=0.35mm, , black]  (4.9,0.5) -- (6.0,0.5);
 	\draw[<-][ line width=0.35mm, , black]  (5.9,0.5) -- (7.0,0.5);
 	\draw[<-][ line width=0.35mm, , black]  (6.9,0.5) -- (8.0,0.5);
 
  	\node at (4.0,.9) {Fig.~(\ref{Fig1_a0.75}) left};

 	\filldraw (0,1.5) circle (2pt);
 	
 	\node at (-1.3,0.5) {$\displaystyle a=\frac{3}{4} \text{, } LRp$};
 	\node at (8.4,0.5) {$Bp$};
 	\filldraw (0,0.5) circle (2pt);
 	\filldraw (8,0.5) circle (2pt);

 	\draw[ line width=0.35mm, , black]  (0.0,-1.5) -- (0.3,-1.5);
 	\draw[<-][ line width=0.35mm, , black]  (0.1,-1.5) -- (1.0,-1.5);
 	\draw[<-][ line width=0.35mm, , black]  (0.9,-1.5) -- (2.0,-1.5);
 	\draw[<-][ line width=0.35mm, , black]  (1.9,-1.5) -- (3.0,-1.5);
 	\draw[<-][ line width=0.35mm, , black]  (2.9,-1.5) -- (4.0,-1.5);
 	\draw[<-][ line width=0.35mm, , black]  (3.9,-1.5) -- (5.0,-1.5);
 	\draw[<-][ line width=0.35mm, , black]  (4.9,-1.5) -- (6.0,-1.5);
 	\draw[<-][ line width=0.35mm, , black]  (5.9,-1.5) -- (7.0,-1.5);
 	\draw[<-][ line width=0.35mm, , black]  (6.9,-1.5) -- (8.0,-1.5);
 	\node at (4.0,-1.1) {Fig.~(\ref{Fig1_a0.25})};
 	\filldraw (0,-0.5) circle (2pt);

 	%
 	\node at (-1.3,-1.5) {$\displaystyle a=\frac{1}{4} \text{, } LRp$};
 	\node at (8.4,-1.5) {$Bp$};
 	\filldraw (0,-1.5) circle (2pt);
 	\filldraw (8,-1.5) circle (2pt);

 	\end{tikzpicture}
 	}
 	\caption{Fixed point stability, obtained via the measurements of the $\beta / \nu$ critical exponent, for $q=1$. The red zone represents a unique fixed point, the Bp one. There is no flow while tuning $r$.
 	 The arrows describe a flow between two fixed points. The red line represents again the Bp fixed point and the blue line is a line of fixed points, the LRp ones.} 
 	\label{Figq1}
 \end{figure}

We can therefore summarize the numerical findings for $q=1$ in Fig.~(\ref{Figq1}).

\subsection{The $q=2$ phase diagram}
\label{sec:q2}
\begin{figure}[!ht]
	\begin{center}
		\includegraphics[width=16cm,height=10cm]{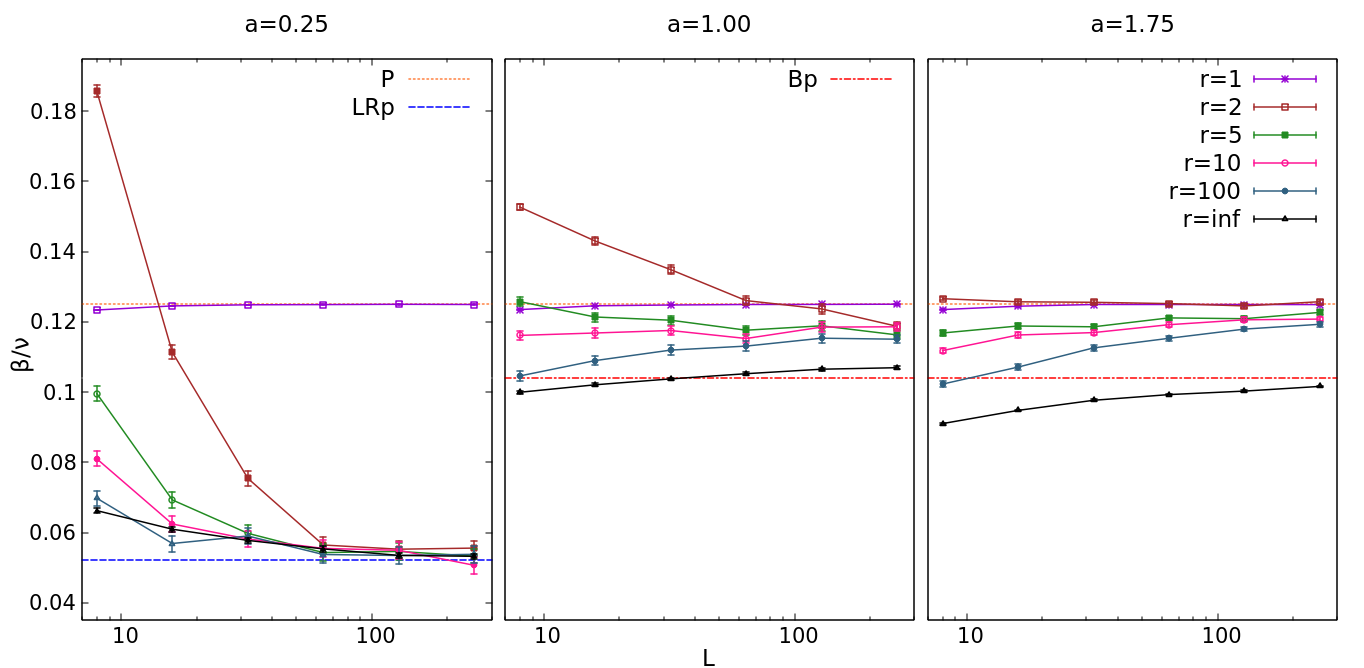}
	\end{center}
	\caption{$q=2$ Potts model with long-range disorder:  $\beta/\nu$ vs. $L$ for $a=0.25$ on the left, $a=1.00$ on the middle and $a=1.75$ on the right.} 
	\label{FigQ2}
\end{figure}

We consider here in detail the $q=2$ case. In the short-range regime $a\geq 2$, the disorder is marginal and the P and SR points coincide.
When $a^*(2)< a< 2$ we can numerically probe the existence of the LR point, predicted in \cite{Rajabpour_2007,dudka2016critical}. Here, 
it will correspond to a fixed point at finite disorder $r^*_{LR}$, $1 < r^*_{LR} < \infty$. We estimate $a^*(2)$ to be in the interval  $1/2<a^*(2)<3/4$, but the determination of its exact value goes beyond the aims of this paper. 
When $a< a^*(2)$ there are just two fixed points, the P and the LRp ($r=\infty$) ones. In this case, the LRp becomes the stable one and we observe a flow from the P point to the LRp point.
These different behaviours are shown in Fig.~(\ref{FigQ2}). In particular: 
\begin{itemize}
\item  In the left panel, we show the measured values of $\beta/\nu (L)$ for $a=1/4$. 
We clearly observe that for all values of disorder, the effective magnetic exponent goes to the same value as the infinite disorder one, $\beta^{LRp}/\nu^{LRp}=5/96\sim 0.0521$, see Table~\ref{LRptable}. 
We obtain a similar behaviour for $a=0.5$, with $\beta/\nu(L) \to 0.086$ in the $L\to\infty$ limit, again very close to the value $\beta^{LRp}/\nu^{LRp}$ reported in Table \ref{LRptable}. 
\item In the middle panel, we show our data for $a=1$ and we observe three different large size limits: i) for $r=1$ the magnetic exponent is the one of the short-range Ising, 
$\beta^{P}/\nu^{P}=1/8=0.125$; ii) for $r=\infty$ the magnetic exponent is close to the value $\beta^{Bp}/\nu^{Bp} =5/48$. As already mentioned for the case of infinite disorder,
see Section~\ref{MCrinfty}, for $a\simeq 1$, it is difficult to distinguish the LRp point from the Bp point; iii) for any finite disorder, there is an unique large size limit,  $\lim_{L\to \infty}\beta/\nu(L)=\beta^{LR}/\nu^{LR} \simeq 0.115-0.120$. This corresponds to the LR fixed point. The curves converge to the curve with a finite disorder $r \simeq 10$, that is the value in which we observe the smallest corrections. 

We observed a similar behaviour for $a=0.75$  and $a=1.25$ with the scaling limit $\lim_{L\to \infty}\beta/\nu(L) \simeq 0.108$ and $\lim_{L\to \infty}\beta/\nu(L) \simeq 0.122$ respectively. 
\item The right panel shows the measured values of $\beta/\nu(L)$ for $a=1.75$. In that case, the measurements for finite disorder also converge to a single value $\beta^{LR}/\nu^{LR}$ which is very close to the P=SR value $\beta^{P}/\nu^{P}=1/8$. For all values of $a \geq 3/2$, due to strong finite size corrections, we find values close to the P=SR values. 
\end{itemize}
%
%

Since, for values of $3/2\leq a<2$, the LR point is close to the P fixed point and we can not really distinguish via the $\beta/\nu$ exponent  these two points, we will then consider another way of checking the existence of the LR point.  We consider the moments of the spin-spin correlation function,
\begin{equation} 
\mathbb{E}[  <s(0) s(x)>^{\displaystyle n}] \simeq x^{\displaystyle-\eta_n} \; .
\label{def:ss}
\end{equation}
and we are interested in extracting the exponents $\eta_n$. While $\eta_1=2\beta/\nu$ gives  access to the same exponent that we computed above via the FK clusters dimensions, $\eta_n$ for $n>1$ are new exponents. 
The authors of \cite{dudka2016critical} were able to compute $\eta_2$  for $ 0.995 \leq a < 2$. They obtained, at the lowest order in disorder, the following:
\begin{align} 
\label{fedoeta2}
\eta_2 = \frac{1}{2} - \frac{(2 - a)}{4} +O((2 - a)^2).
\end{align}
In this computation the lower range, $a=0.995$, is obtained as a limit of the stability of their perturbative fixed point at the second order. 

\begin{figure}[!ht]
	\begin{center}
	\includegraphics[width=12cm,height=10cm]{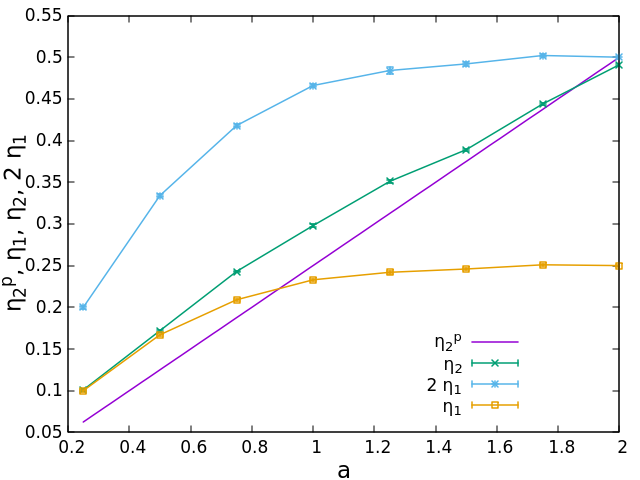}
	\end{center}
	\caption{$q=2$ Potts model with long-range disorder:  $\eta_1, \eta_2$ and $\eta_2^p$ vs $a$.} 
	\label{Figeta}
\end{figure}

\begin{figure}[!ht]
	\hspace*{1.2cm}
	\scalebox{1.}{%
		\begin{tikzpicture}[scale=1.00]

		\draw[ thick]  [->](0.0,-2.5) -- (9.0,-2.5);
		\draw[ thick]  [->] (0.0,-2.5) -- (0.0,8.5);
		
		\node at (3.9,8.9) {$q=2$};
		\node at (-0.5,8.5) {$a$};
		\node at (9.3,-2.5) {$1/r$};
		\node at (8.,-3.) {$1$};

		\draw [line width=0.85mm, magenta!90] plot [smooth, tension=0.9] coordinates {(8,5.5) (6.1,3.4) (0,0.2)};
		
		\filldraw (.65,.5) circle (2pt); 
		
		\node at (2.5,1.8) {$LR$};
		\filldraw (2.69,1.51) circle (2pt); 
		
		\filldraw (4.615,2.5) circle (2pt); 
%
		\filldraw (6.25,3.5) circle (2pt);
		
		\node at (7.1,4.8) {$LR$};
		\filldraw (7.45,4.5) circle (2pt);
		
		\draw[ line width=0.85mm, , orange!80]  (8.0,-2.5) -- (8.0,7.5);
		\draw[ line width=0.85mm, , red!80]  (0.0,3.5) -- (0.0,7.5);
		\filldraw (8.,-2.5) circle (2pt);
		
		\draw[ line width=0.85mm, , blue!80]  (0.0,3.5) -- (0.0,-2.5);
		\filldraw (0.,-2.5) circle (2pt);
		\filldraw (0.,3.5) circle (2pt);
		\node at (-.6,0.2) {$a^*(2)$};
		\filldraw (0.0,0.2) circle (2pt); 
	

		

%


		\draw[ dashed ]  (0.0,5.5) -- (8.0,5.5); 
		\node at (-0.7,5.5) {$a=2$};
		\filldraw (0.,5.5) circle (2pt);
		\filldraw (8.,5.5) circle (2pt);

		\node at (4.0,4.8) {Fig.(\ref{FigQ2}) right};
		\draw[ line width=0.35mm, , black]  (0.0,4.5) -- (0.3,4.5);
		\draw[->][ line width=0.35mm, , black]  (0.1,4.5) -- (1.0,4.5);
		\draw[->][ line width=0.35mm, , black]  (0.9,4.5) -- (2.0,4.5);
		\draw[->][ line width=0.35mm, , black]  (1.9,4.5) -- (3.0,4.5);
		\draw[->][ line width=0.35mm, , black]  (2.9,4.5) -- (4.0,4.5);
		\draw[->][ line width=0.35mm, , black]  (3.9,4.5) -- (5.0,4.5);
		\draw[->][ line width=0.35mm, , black]  (4.9,4.5) -- (6.0,4.5);
		\draw[->][ line width=0.35mm, , black]  (5.9,4.5) -- (7.0,4.5);
		\draw[-][ line width=0.35mm, , black]  (6.9,4.5) -- (7.9,4.5);
		\draw[<-][ line width=0.35mm, , black]  (7.8,4.5) -- (8,4.5);

		\node at (-1.1,4.5) {$a=\displaystyle\frac{7}{4} \text{, }Bp$};
		\filldraw (0.,4.5) circle (2pt);
		\node at (8.4,4.5) {$P$};
		\filldraw (8.,4.5) circle (2pt);
		

		\node at (-1.8,3.5) {$\displaystyle a=\frac{3}{2} \text{, } LRp=Bp$};
		
		
		
		%
		%
		\filldraw (0,2.5) circle (2pt);

		\draw[ line width=0.35mm, , black]  (0.0,1.5) -- (0.3,1.5);
		\draw[->][ line width=0.35mm, , black]  (0.1,1.5) -- (1.0,1.5);
		\draw[->][ line width=0.35mm, , black]  (0.9,1.5) -- (2.0,1.5);
		\draw[-][ line width=0.35mm, , black]  (1.9,1.5) -- (3.0,1.5);
		\draw[<-][ line width=0.35mm, , black]  (2.9,1.5) -- (4.0,1.5);
		\draw[<-][ line width=0.35mm, , black]  (3.9,1.5) -- (5.0,1.5);
		\draw[<-][ line width=0.35mm, , black]  (4.9,1.5) -- (6.0,1.5);
		\draw[<-][ line width=0.35mm, , black]  (5.9,1.5) -- (7.0,1.5);
		\draw[<-][ line width=0.35mm, , black]  (6.9,1.5) -- (8.0,1.5);

		\node at (-1.3,1.5) {$a = \displaystyle 1 \text{, }LRp$};
		\node at (8.4,1.5) {$P$};
		\filldraw (8,1.5) circle (2pt);
		\filldraw (0,1.5) circle (2pt);
		\node at (5.2,1.8) {Fig.(\ref{FigQ2}) middle};
		
		\filldraw (0,0.5) circle (2pt);

		\draw[ line width=0.35mm, , black]  (0.0,-1.5) -- (0.3,-1.5);
		\draw[<-][ line width=0.35mm, , black]  (0.1,-1.5) -- (1.0,-1.5);
		\draw[<-][ line width=0.35mm, , black]  (0.9,-1.5) -- (2.0,-1.5);
		\draw[<-][ line width=0.35mm, , black]  (1.9,-1.5) -- (3.0,-1.5);
		\draw[<-][ line width=0.35mm, , black]  (2.9,-1.5) -- (4.0,-1.5);
		\draw[<-][ line width=0.35mm, , black]  (3.9,-1.5) -- (5.0,-1.5);
		\draw[<-][ line width=0.35mm, , black]  (4.9,-1.5) -- (6.0,-1.5);
		\draw[<-][ line width=0.35mm, , black]  (5.9,-1.5) -- (7.0,-1.5);
		\draw[<-][ line width=0.35mm, , black]  (6.9,-1.5) -- (8.0,-1.5);
		\node at (4.0,-1.1) {Fig.(\ref{FigQ2}) left};
		\filldraw (0,-0.5) circle (2pt);

		\node at (-1.3,-1.5) {$a= \displaystyle \frac{1}{4} \text{, }LRp$};
		\node at (8.4,-1.5) {$P$};
		\filldraw (0,-1.5) circle (2pt);
		\filldraw (8,-1.5) circle (2pt);

		\end{tikzpicture}
	}
	\caption{Fixed point stability, obtained via the measurements of the $\beta/ \nu $ critical expo-
		nent, for $q=2$. The dashed line represents the crossover with the purely SR physics at $a=2$. The arrows describe a flow between two fixed points. The orange line represents the P (unique) fixed point, the red one the Bp (unique) fixed point, while the blue and magenta lines are lines of fixed points, respectively the LRp and the LR ones. $a^*(2)$ indicates the crossover between the LR and LRp points and we located it slightly below $a=0.75$. }
	\label{Figq2}
\end{figure}
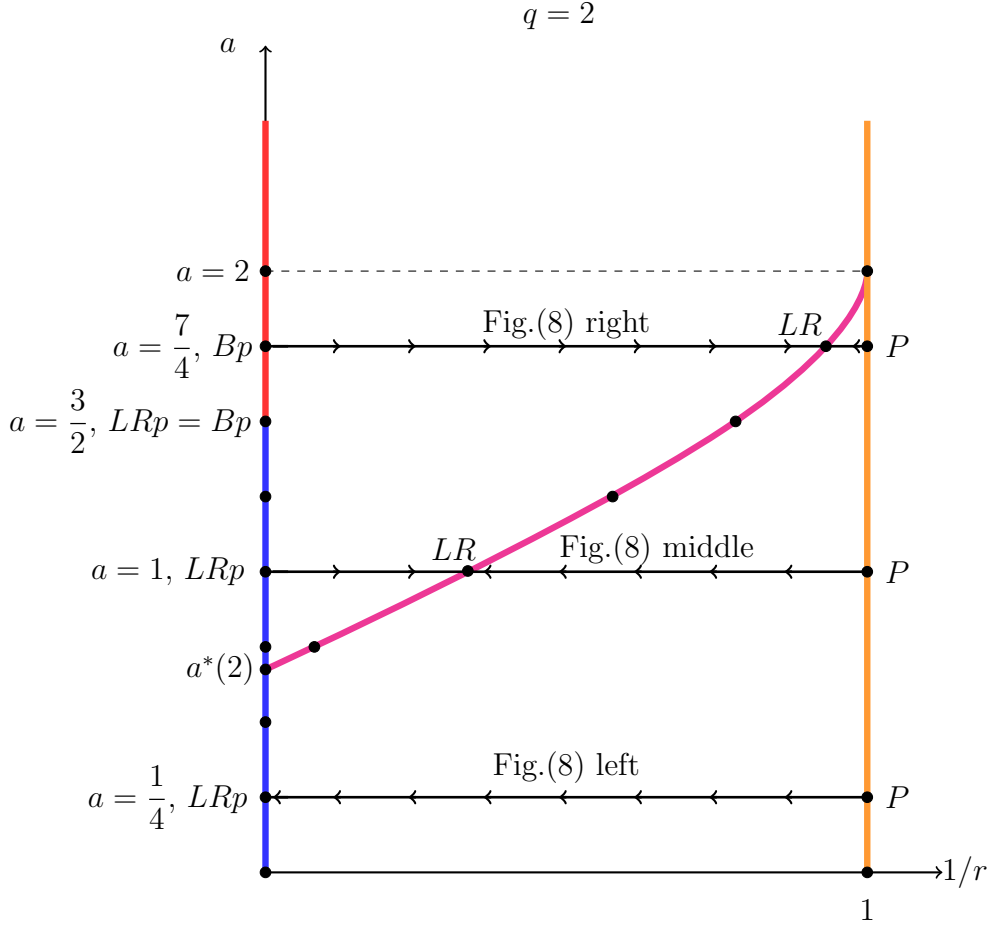

 In Table~\ref{eta}, we report our measured values of $\eta_1$ and $\eta_2$ for various values of $a$. Note that the values for $\eta_1$ can be compared to the values of $\beta/\nu$ previously obtained. One can verify that the two ways of measuring the same exponent give totally consistent results.  
For each value $a$, the measurements have been done for a fixed value of disorder $r$ chosen in order to minimise corrections to  scaling in the measurement of the fractal dimension. 
The last column of the table contains the prediction of Eq.~(\ref{fedoeta2}) at the first order, $\eta_2^p$.
\begin{table}[h]
\centering
\renewcommand{\arraystretch}{1.5}
\footnotesize

\begin{tabular}{|p{.9cm}|P{1.9cm}|P{1.2cm}|P{.9cm}|P{.9cm}|}
\hline
$a$   & $\eta_1 (=2\beta/\nu)$ &  $\eta_2$  &  $r$ & $\eta_2^p$ \\
\hline 
 0.25  &  0.100    &   0.101  &  100 &  0.0625  \\
 \hline
 0.50  & 0.167       & 0.172   &  100 &  0.125 \\
 \hline
 0.75  & 0.209         & 0.243  &  10 & 0.1875 \\
 \hline
 1.00  & 0.233        & 0.298(2)   & 10 & 0.25\\
 \hline
1.25  & 0.242(2)   & 0.351(2)      & 5  & 0.3125 \\
\hline
1.50  & 0.246       & 0.389    & 5 & 0.375 \\
\hline
1.75  & 0.251    & 0.444    & 2 & 0.4375  \\
\hline
2.00  &  0.250   &  0.491  &   1  &  0.50 \\
\hline
\end{tabular}
\caption{$\eta_1$ and $\eta_2$ measured with the disorder $r$ for $0.25 \leq a < 2$. $\eta_2^p= 1/2 - (2 - a)/4$ is the predicted value of Eq.~(\ref{fedoeta2}) at the first order.
Errors bars on the measured values $\eta_1$ and $\eta_2$ are shown in parenthesis and are smaller than one on the last digit otherwise.}
\label{eta}
\end{table}
Our results are also shown in Fig.~(\ref{Figeta}). 
The agreement is good between the measured $\eta_2$ and the predicted one $\eta_2^p$ for $1.5 \leq a \leq 2.0$. For smaller values of $a$, 
the deviation is larger but also the second order correction $\simeq (a-2)^2$. Quite interestingly, for $a \leq 0.5$, $\eta_1 \simeq \eta_2$.  For the $r=\infty$ disorder point, we recall that the random FK clusters (and the spin variables) are completely determined by the disorder configurations and the thermal fluctuations are frozen, see Section~\ref{sec:rinfty}. This implies that the distribution function $\mathcal{S}(p_{12}(x))$ of the probability $p_{12}$ that the spin $s(0)$ and $s(x)$ belong to the same FK cluster, has the form  $\mathcal{S}(y)=a_0\delta_{y,0}+a_1\delta_{y,1}$, with $a_0+a_1=1$. This in turn implies $\eta_1=\eta_2$. So the fact that, at finite $r$, where the spin degrees of freedom are not frozen, we obtain $\eta_1\sim\eta_2$ is a very non-trivial test of the fact that the infinite disorder point describes the physics of the system for these values of $a$.  
%
The conclusions of our analysis are summarised in  Fig.~(\ref{Figq2}).

\subsection{The $q=3$ phase diagram}
In the short-range regime $a>2$, the disorder is now relevant and the critical properties are described by the SR, which is located at finite $r^*_{SR}$,  $r^*_{SR}= 6.08 (12)$ \cite{picco2022}. The two fixed points, SR for $ 1<r<\infty$ and Bp for $r = \infty$ are observed for all values of $q$ up to infinite \cite{picco1997weak,cardy1997critical,jacobsen2000large}. The $r=1$ point corresponds to the P point for $q\in [1,4]$ and to a first order phase transition for $q>4$. 

\begin{figure}[!ht]
	\begin{center}
		\includegraphics[width=16cm,height=10cm]{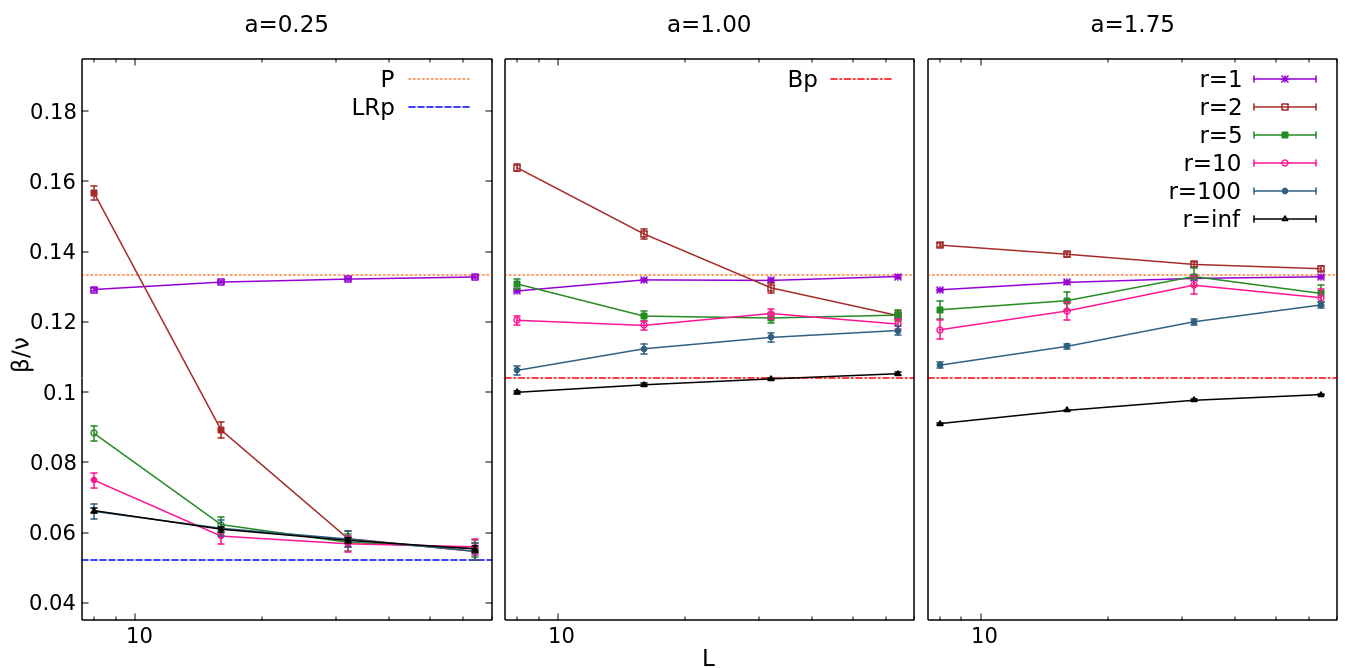}
	\end{center}
	\caption{$q=3$ Potts model with long-range disorder:  $\beta/\nu$ vs. $L$ for $a=0.25$ on the left, $a=1.00$ on the middle and $a=2.00$ on the right.} 
	\label{FigQ3}
\end{figure}

When $a<2$, the long-range disorder is relevant for  $a<2/\nu^{SR}\sim 1.96 $. We show below that our numerical simulations support the existence of a stable  LR that is located at finite $1<r^{*}_{SR}<r^{*}_{LR}$.
We observed, moreover, that the LR point tends to merge with the LRp one, located at a critical value $a^*(3)$ slightly smaller than $0.75$. 

\begin{figure}[!ht]
	\hspace*{1.2cm}
		\scalebox{1.}{%
			\begin{tikzpicture}[scale=1.00]

			\draw[ thick]  [->](0.0,-2.5) -- (9.0,-2.5);
			\draw[ thick]  [->] (0.0,-2.5) -- (0.0,8.5);
			
			\node at (3.9,8.9) {$q=3$};
			\node at (-0.5,8.5) {$a$};
			\node at (9.3,-2.5) {$1/r$};
			\node at (8.,-3.) {$1$};

			\draw [line width=0.85mm, magenta!90] plot [smooth, tension=1] coordinates {(6,5.2)  (3,2) (0,0.2)};
			
			\draw[ line width=0.85mm, , cyan!80]  (5.97,5.2) -- (5.97,7.5);
			
			\filldraw (5.97,5.2) circle (2pt);
%
%
%

			\filldraw (.7,.5) circle (2pt);
			
			\node at (1.9,1.8) {$LR$};
			\filldraw (2.35,1.5) circle (2pt); 
			
			\filldraw (3.57,2.5) circle (2pt); 
			
			\filldraw (4.56,3.5) circle (2pt);
			
			\node at (5.1,4.8) {$LR$};
			\filldraw (5.45,4.5) circle (2pt);

			\draw[ dashed ]  (0.0,5.2) -- (8.0,5.2);
			\node at (4.9,5.5) {$LR=SR$};

			\draw[ line width=0.85mm, , orange!80]  (8.0,-2.5) -- (8.0,7.5);
			\draw[ line width=0.85mm, , red!80]  (0.0,3.5) -- (0.0,7.5);
			\filldraw (8.,-2.5) circle (2pt);
			
			\draw[ line width=0.85mm, , blue!80]  (0.0,3.5) -- (0.0,-2.5);
			\filldraw (0.,-2.5) circle (2pt);
			\filldraw (0.,3.5) circle (2pt);
			\filldraw (0.0,0.2) circle (2pt);

			\node at (-0.9,5.2) {$a =1.96 $};
			\filldraw (0.,5.2) circle (2pt);
			\filldraw (8.,5.2) circle (2pt);
%
			
			\node at (3.4,4.1) {Fig.~(\ref{FigQ3}) right};
			\node at (-1.1,4.5) {$a=\displaystyle \frac{7}{4}\text{, }Bp$};
			\filldraw (0.,4.5) circle (2pt);
			\node at (8.4,4.5) {$P$};
			\filldraw (8.,4.5) circle (2pt);
			
					\draw[ line width=0.35mm, , black]  (0.0,4.5) -- (0.3,4.5);
					\draw[->][ line width=0.35mm, , black]  (0.1,4.5) -- (1.0,4.5);
					\draw[->][ line width=0.35mm, , black]  (0.9,4.5) -- (2.0,4.5);
					\draw[->][ line width=0.35mm, , black]  (1.9,4.5) -- (3.0,4.5);
					\draw[->][ line width=0.35mm, , black]  (2.9,4.5) -- (4.0,4.5);
					\draw[->][ line width=0.35mm, , black]  (3.9,4.5) -- (5.0,4.5);
					\draw[-][ line width=0.35mm, , black]  (4.9,4.5) -- (6.0,4.5);
					\draw[<-][ line width=0.35mm, , black]  (5.9,4.5) -- (7.0,4.5);
					\draw[<-][ line width=0.35mm, , black]  (6.9,4.5) -- (7.9,4.5);
					\draw[<-][ line width=0.35mm, , black]  (7.8,4.5) -- (8,4.5);

			\node at (-1.8,3.5) {$\displaystyle a=\frac{3}{2} \text{, } LRp=Bp$};
			
			
			

			\filldraw (0,2.5) circle (2pt);

			\draw[ line width=0.35mm, , black]  (0.0,1.5) -- (0.3,1.5);
			\draw[->][ line width=0.35mm, , black]  (0.1,1.5) -- (1.0,1.5);
			\draw[-][ line width=0.35mm, , black]  (0.9,1.5) -- (2.0,1.5);
			\draw[-][ line width=0.35mm, , black]  (1.9,1.5) -- (3.0,1.5);
			\draw[<-][ line width=0.35mm, , black]  (2.9,1.5) -- (4.0,1.5);
			\draw[<-][ line width=0.35mm, , black]  (3.9,1.5) -- (5.0,1.5);
			\draw[<-][ line width=0.35mm, , black]  (4.9,1.5) -- (6.0,1.5);
			\draw[<-][ line width=0.35mm, , black]  (5.9,1.5) -- (7.0,1.5);
			\draw[<-][ line width=0.35mm, , black]  (6.9,1.5) -- (8.0,1.5);
			
			\node at (4.5,1.8) {Fig.~(\ref{FigQ3}) middle};
			\node at (-1.3,1.5) {$a=\displaystyle 1 \text{, }LRp$};
			\node at (8.4,1.5) {$P$};
			\filldraw (0,1.5) circle (2pt);
			\filldraw (8,1.5) circle (2pt);
			
			\node at (-0.6,0.2) {$a^*(3)$};
			\filldraw (0,0.5) circle (2pt);

			\filldraw (0,-0.5) circle (2pt);

			\draw[ line width=0.35mm, , black]  (0.0,-1.5) -- (0.3,-1.5);
			\draw[<-][ line width=0.35mm, , black]  (0.1,-1.5) -- (1.0,-1.5);
			\draw[<-][ line width=0.35mm, , black]  (0.9,-1.5) -- (2.0,-1.5);
			\draw[<-][ line width=0.35mm, , black]  (1.9,-1.5) -- (3.0,-1.5);
			\draw[<-][ line width=0.35mm, , black]  (2.9,-1.5) -- (4.0,-1.5);
			\draw[<-][ line width=0.35mm, , black]  (3.9,-1.5) -- (5.0,-1.5);
			\draw[<-][ line width=0.35mm, , black]  (4.9,-1.5) -- (6.0,-1.5);
			\draw[<-][ line width=0.35mm, , black]  (5.9,-1.5) -- (7.0,-1.5);
			\draw[<-][ line width=0.35mm, , black]  (6.9,-1.5) -- (8.0,-1.5);
			\node at (-1.3,-1.5) {$a=\displaystyle \frac{1}{4}\text{, }LRp$};
			\node at (8.4,-1.5) {$P$};
			\filldraw (0,-1.5) circle (2pt);
			\filldraw (8,-1.5) circle (2pt);
			\node at (4.0,-1.1) {Fig.~(\ref{FigQ3}) left};

			\end{tikzpicture}
		}
	\caption{Fixed point stability, obtained via the measurements of the $\beta/ \nu $ critical expo-
		nent, for $q=3$. The dashed line represents the crossover with the purely SR physics at $a=1.96$. The arrows describe a flow between two fixed points. The orange line represents the P (unique) fixed point, the red line the Bp (unique) fixed point and the cyan one the SR (unique) fixed point.  The blue and magenta lines are lines of fixed points, respectively the LRp and the LR ones. $a^*(3)$ represents the crossover between the LR and LRp points and we located it slightly below $a=0.75$.}
	\label{Figq3}
\end{figure}
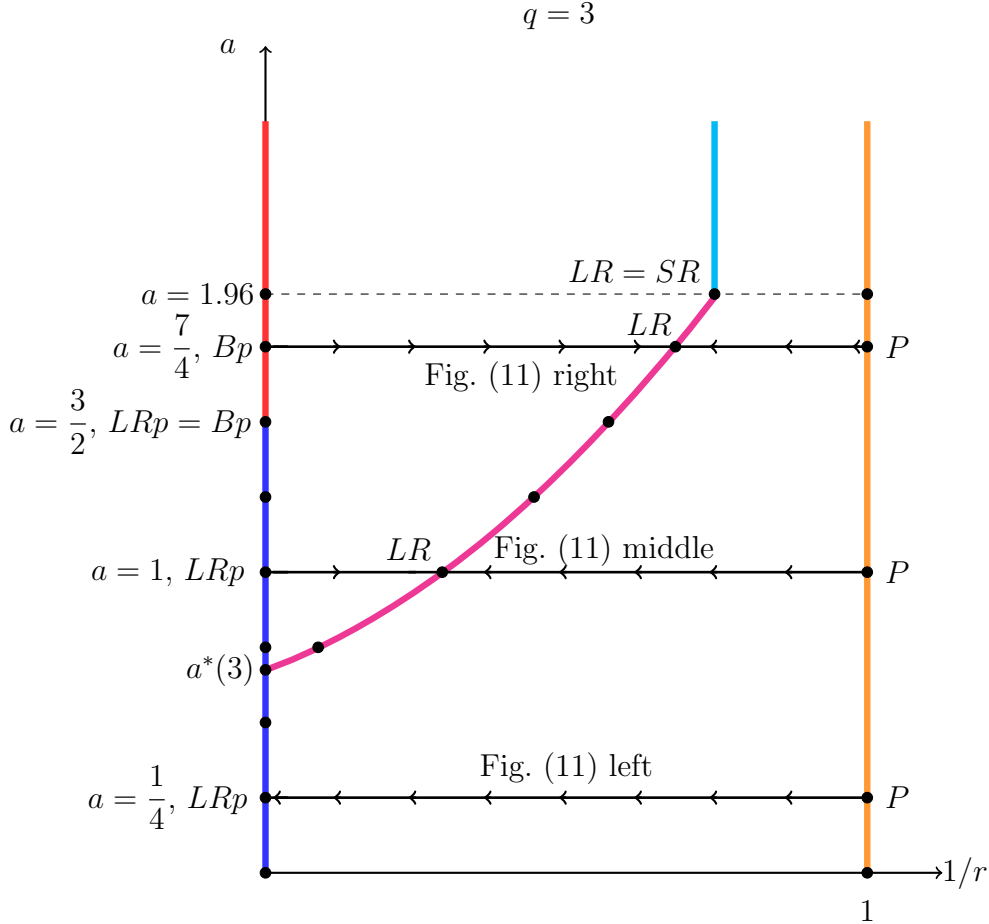

Our numerical findings for $q=3$ are shown in Fig.~(\ref{FigQ3}). Again, we obtain three different behaviours:~
\begin{itemize}
\item In the left panel, we show $\beta/\nu$ for $a=1/4$. For all values of disorder $r>1$, in the large size limit, the effective magnetic exponent 
goes to the value $\beta^{LRp}/\nu^{LRp}=5/96$ while it goes to the value of the pure $q=3$-Potts model $\beta^P/\nu^P= 2/15$ for $r=1$. 
We observe a similar behaviour for $a=0.5$ with $\beta^{LRp}/\nu^{LRp}\simeq 0.09$. 

\item In the middle panel, we show results for $a=1$, with three different large size limits: i) for $r=1$, the magnetic exponent is the one of the pure $q=3$-Potts model $\beta^P/\nu^P= 2/15$; ii) for $r=\infty$, the magnetic exponent goes towards a value $\simeq 5/48$. Notice that this is the same as the one mention in the section above for $q=2$ and $a=1$: we recall that at this point the exponents do not depend on $q$; iii) for any finite disorder, a unique limit is obtained with $\beta^{LR}/\nu^{LR} \sim 0.12$ 
corresponding to a LR fixed point with finite disorder $r \simeq 10$. 

A similar behaviour is observed for $a=0.75, 1.25$ with $\beta^{LR}/\nu^{LR} \sim 0.11$ and $\beta^{LR}/\nu^{LR} \sim 0.13$ respectively.

\item In the right panel, we show results for $a=1.75$ .
For $a=1.50$ and $a=1.75$, $\beta/\nu$ is compatible, at large distances, with either with the P and the SR values, reported in Table \ref{Ptable} and in Table \ref{SRtable}, which are very close. Our measurements therefore do not allow to discriminate between these two values, but according to the arguments discussed in Section~\ref{CD}, we conjecture the LR point will merge with the SR point at $a\sim1.96$, as illustrated in Fig.~(\ref{Figq3}). 

\end{itemize}

In conclusion, for $a < a^*(3) \simeq 3/4$, we find that the system flows to the LRp point, while for $a^*(3) < a <1.96$, we find the LR at some intermediate value of disorder with $\beta^{Bp}/\nu^{Bp}<\beta^{LR}/\nu^{LR}<\beta^{SR}/\nu^{SR}$. 

Our findings are summarized  in Fig.~(\ref{Figq3}).

\subsection{Thermal behaviour for the $q=3$-Potts model with long-range correlated disorder}
\label{sec:nu}
In this section, to test Eq.~\ref{nulr}, we present results for the measurement of the exponent $\nu$ as a function of $a$. A convenient way to measure it is to consider the FK cluster wrapping probability 
\begin{equation}
\label{def:wp}
\mathcal{P}_w(r,J_1,L)=\text{Prob. of having a wrapping FK cluster on a lattice of size $L$}.
\end{equation} 
This is a quantity which is constant at the critical point 
for a fixed value $r= J_2/J_1$. 
$J_1$ is coupled to the thermal behaviour and we expect:
\begin{equation}
\label{eq:wrscaling}
\begin{split}
\mathcal{P}_w(r,J_1 (1+\epsilon),L) & = f( \epsilon L^{1/\nu}) \\
 & \simeq f(0) + f'(0) \epsilon L^{1/\nu} + \frac{f''(0)}{2} \epsilon^2 L^{2/\nu}  + \frac{f'''(0)}{3!} \epsilon^3 L^{3/\nu} + \cdots 
\end{split}
\end{equation}

for a small $\epsilon$, which measures the deviation from the critical point.   $\mathcal{P}_w(r,J_1,L)$ should also depend on $r$ and one expect also corrections to scaling. 
These dependencies can be ignored if we consider the derivative 
\begin{equation}
\begin{split}
\mathcal{P}_w'(\epsilon) & = \frac{\mathcal{P}_w(r,J_1 (1+\epsilon),L) - \mathcal{P}_w(r,J_1 (1-\epsilon),L)}{2 \epsilon} \\
& \simeq f'(0) L^{1/\nu} + \frac{f'''(0)}{3!} \epsilon^2 L^{3/\nu} + \cdots \; .
\label{eq:wrscaling2}
\end{split}
\end{equation}
We first attempted to compute $1/\nu$ by considering $\mathcal{P}_w'$ at the first order, but the results were not convergent for large $L$. 
This is due to the fact that the errors on the measurement grow like the inverse of $\epsilon$. While averaging over one million samples of disorder configurations, 
we found that errors bars allow us to decrease $\epsilon$ only down to $0.01$. For this value, the second correction can not be neglected. We then considered 

\begin{equation} 
\tilde{\mathcal{P}'_w} (\epsilon)
= \frac{4 \mathcal{P}'_w(\epsilon)  - \mathcal{P}'_w(2\epsilon)}{3} = f'(0) L^{1/\nu} (1 + O( \epsilon^4 L^{4/\nu}))
\label{ddw}
\end{equation}

\begin{figure}[!ht]
\begin{center}
\includegraphics[width=12cm,height=10cm]{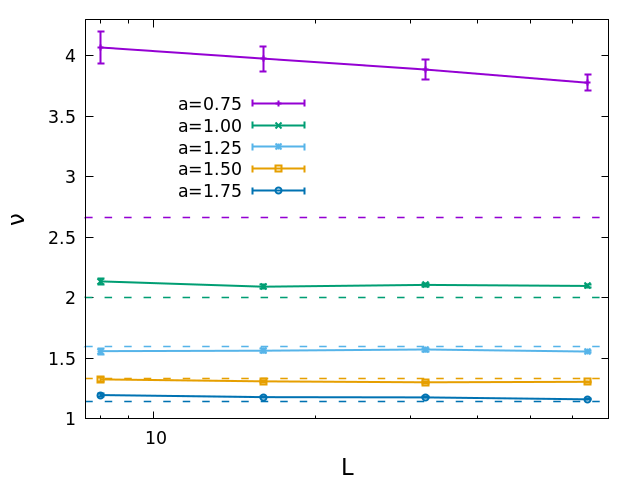}
\end{center}
\caption{$\nu$ vs. $L$ for the various values of $a$ shown in the caption.}
\label{FigQ3_nu}
\end{figure}
In Fig.~(\ref{FigQ3_nu}), we show the results for the effective $\nu$ as a function of $L$ obtained from a two point fit of our data to the form of Eq.~(\ref{ddw}) for $a \geq 0.75$ 
and also the value $2/a$ as a dotted line. For each value of $a$, we choose the value of $r$ which seems closer to the critical point with the least 
finite size corrections. For $a \geq 1.25$, we used 
the data for $r=5$, while for $a=0.75$ and $1.00$, we used the data for $r=10$.  Note that for $a=0.75$, we still have large error bars even if for this value 
of $a$, we averaged over $10$ millions samples of disorder configurations (the same statistics is used for $a=1.00$, with much smaller error bars, for $a> 1$, 
we averaged over one million samples). For $a=0.75$, we also observe in Fig.~(\ref{FigQ3_nu}) that there are strong finite size corrections. For smaller values of $a$, we 
were not able to obtain convergent results. Note that this corresponds to values of $a$ for which we expect that the critical point is at an infinite disorder. 

In conclusion, we observe that the value of $\nu$ increase slowly as one decreases $a$. For $1 \leq a < 2$, $\nu$ is close to the prediction in Eq.~(\ref{nulr}). For $a=0.75$, it seems 
to be bigger than expected with $\nu \simeq 3.5$, but in this case, we observe strong finite size corrections. For smaller value of $a$, we can not measure $\nu$. 

\subsection{The effects of higher disorder cumulants}
\label{sec:hc}
Let us come back to the value $q=1$ where the disordered Potts model becomes a long-range percolation model. 

Without passing via a Potts model, a long-range percolation model can be directly defined by using the level sets of a fractional Gaussian free field (fGFF) with negative Hurst exponent $H=-a/2$ \cite{stanley92, Schmittbuhl_1993,Janke_2017, de_Castro_2018,Javerzat_2020}. 

\begin{figure}[!ht]
	\begin{center}
		\includegraphics[width=12cm,height=10cm]{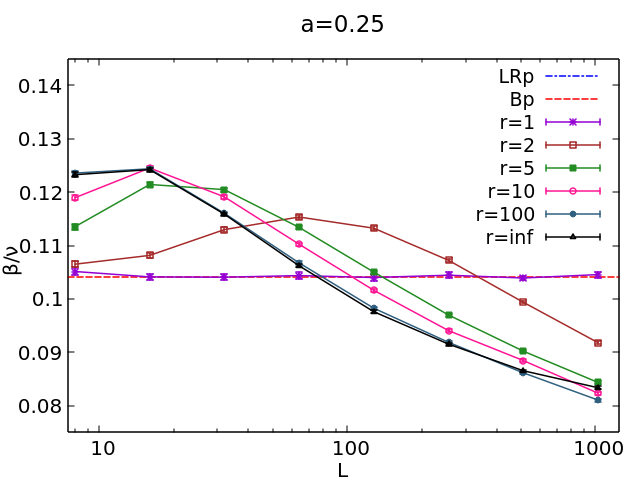}
		\end{center}
	\caption{$\beta/\nu$ vs. $L$ at $q=1$ with $a=0.25$ and disorder generated by using fGFF, see Appendix~\ref{fgff}.} 
	\label{FigIVSG}
\end{figure}

In our approach, we can use the fGFF to generate a long-range disorder distribution, see Appendix~\ref{fgff}. In doing so for the $q=1$ Potts model, we obtain a one-parameter ($r$) family of Gaussian long-range percolation models. In our simulations, we have verified that, analogously to what explained in Section~\ref{sec:q1}, all these models flow to the point $r=\infty$, see Fig.~(\ref{FigIVSG}). At this point, our long-range percolation models are expected to have the same critical properties as the models studied in \cite{stanley92, Schmittbuhl_1993,Janke_2017, de_Castro_2018,Javerzat_2020}. More specifically, as we used toric boundary conditions, our model coincides at $r=\infty$ to the one studied in \cite{Javerzat_2020}. 

Our approach allows then to compare directly the LRp points obtained using different distributions, in particular non-Gaussian ones. The disorder distribution generated by the  $n$-replicated Ising model is an instance of such distributions. In the following, we use the suffix $I$ ($G$) for the disorder distribution generated by $n$-replicated Ising model (fractional Gaussian free field). 

For the two distributions, the Gaussian one based on the fGFF and the non-Gaussian one based on the Ising model, Eq.~(\ref{nulr}) gives the same 
percolation thermal exponent, $\nu=2/a=8$. 

The situation however is different for the exponent $\beta$. 
Let us discuss more in detail the case $a=1/4$, that we have analyzed more carefully in our simulations. 
In Table~\ref{LRptable2} are reported the $r=\infty$ results of our Monte-Carlo measures of $\left(\beta^{LRp}/\nu^{LRp}\right)_I$, obtained by using the $n$-replicated Ising distribution, see Appendix~\ref{nising}, and of $\left(\beta^{LRp}/\nu^{LRp}\right)_G$ obtained with the fGFF distribution, see Appendix \ref{fgff}. In the same table we have also put the measures obtained in  \cite{Janke_2017} that can be then directly compared. 


\begin{table}[!ht]
	\centering
\renewcommand{\arraystretch}{1.2}
\footnotesize
\centering

\begin{tabular}{|P{.7cm}|P{2.4cm}|P{2.4cm}|P{3.4cm}|}
	\hline
$a$ \hspace{0.5cm}& $\left(\beta^{LRp}/\nu^{LRp}\right)_I$ & $\left(\beta^{LRp}/\nu^{LRp}\right)_G$ &  $\left(\beta^{LRp}/\nu^{LRp}\right)_G$ \cite{Janke_2017}  \\ & &  &\\
\hline 
$0.25$ & $0.0522 (4)$ &0.0721(9) &0.0640(4)\\& &  & \\
\hline 
\end{tabular}
\caption{Comparison of $\beta^{LRp}/ \nu^{LRp}$ for Gaussian and non-Gaussian disorders.}
\label{LRptable2}

\end{table} 
The exponents $\left(\beta^{LRp}/\nu^{LRp}\right)_G$ and $\left(\beta^{LRp}/\nu^{LRp}\right)_I$ obtained respectively with the fGFF and 
with the Ising model seems different.  Note that the numerical result $\left(\beta^{LRp}/\nu^{LRp}\right)_I$ agrees perfectly with the Ising model: 
as argued earlier, for the non-Gaussian disorders with $n=1$-Ising copies, the FK clusters have the same fractal dimensions as the Ising spin clusters.

The above results can be explained by noticing that, differently from the short-range disorder, in a long-range disorder, the terms generated by the higher cumulants of the distribution can be relevant. We can conclude that the higher cumulants do not change the existence and the stability of an LRp point but can modify certain critical exponents. 

As far the exponent $\beta^{LRp}/\nu^{LRp}$ is concerned, one can doubt that the differences seen in the simulations are just due to the numerical precision. This is particularly true when $a$ is increasing, see Table~\ref{LRptable}.
However, we can mention the measure of another universal property which is clearly different between the (LRp)$_G$ and the  (LRp)$_I$ and therefore supports the conclusion above.
Indeed in \cite{Javerzat_2020} and in \cite{Picco_2022} the torus two-point connectivity $p_{12}^{G}(r)$ and $p_{12}^{I}(r)$ of, respectively, the fGFF level sets and the Ising clusters were considered.
It was argued that:
\begin{align}
p_{12}^{G}(r)&=  a^{ G}_0 \;r^{ \displaystyle -2  \left(\frac{\beta^{LRp}}{\nu^{LRp}}\right)_G}\left  [1+c^{G}\left(\frac{r}{L}\right)^{\displaystyle 1.875}+o\left(\left(\frac{r}{L}\right)^{\displaystyle 4}\right)\right]\\
p_{12}^{I}(r)&=a^{I}_0\;r^{\displaystyle -2 \left(\displaystyle \frac{ \beta^{LRp}}{\nu^{LRp}}\right)_I}\times \left[1+c^{I}\left(\frac{r}{L}\right)+o\left(\frac{r}{L}\right)\right],
\label{p12}
\end{align}    
where $a_0^{G}$ and $a^{I}_0$ are non-universal constants. The $c^{G}$ and $c^{I}$ are universal quantities that depend on the ratio $L_h/L_v$ between the horizontal and the vertical torus size and on the structure constants of the eventual CFT behind, see \cite{Javerzat_p12}. The $c^{I}$ is known and it has been computed in  
 \cite{Picco_2022} while $c^{G}$, investigated in \cite{Javerzat_2020}, is not known exactly.
 
The exponent $x$ of the sub-leading term $(r/L)^x$ is computed on the assumption that exists a local CFT describing the critical point. This gives  \cite{Javerzat_p12}:
\begin{equation}
\label{nu}
x=2-\frac{1}{\nu}.
\end{equation} 
The exponents in Eq.~(\ref{p12}) are then obtained by using $\nu=1/8$ ($x=1.875$) for the Gaussian distribution and $\nu=1$ ($x=1$) for the Ising one. For the Ising case, where the existence of a CFT behind is well established, the form for $p_{12}^{I}$ has to be considered exact. For the fGFF, the CFT predictions (\ref{p12}), with $\nu=2/a$, were tested numerically only for values of $a>1$ \cite{Javerzat_2020}. For $p_{12}^{G}$ we can extend the results of \cite{Javerzat_2020} to $a=1/4$, by assuming an   analytical $a-$dependence on the critical exponents for all values of $a<3/2$. 
So one can see that the sub-leading term exponents depend strongly on the disorder distribution.

\section{Conclusions}
    
We considered the long-range disordered two-dimensional $q$-Potts model. The disorder is implemented by a bimodal distribution whose long-range nature is induced by auxiliary $\{\sigma_i\}$ spin degrees of freedom to which the Potts model is coupled, see Eq.~(\ref{bimodal}). We have considered different $\{\sigma_i\}$ distributions, the Ising distribution described in Appendix~\ref{nising} and the fGFF distribution described in Appendix~\ref{fgff}. These distributions have the same first cumulant, see Eq.~(\ref{1cum}), and a second cumulant with the same long distance power-law decay, see Eq.(\ref{defa}).   

By Monte-Carlo techniques, we studied the fractal dimension of the $q=1,2,3$-Potts FK clusters for different values of the power-law exponent $a$. The measures were taken at the self-dual point, see  Eq.~(\ref{dual}),  for different values of the strength disorder $r$. The outcome of these analysis are resumed in the phase diagram Fig.~(\ref{ExtHarrfig}), which represents the main finding of this work.

We considered first the infinite disorder ($r=\infty$) fixed point LRp, at which the thermal fluctuations of the Potts degrees of freedom are frozen, and the disorder averages coincide with the ones of a long-range bond percolation model. The LRp critical exponents do not depend on $q$. The measured value of $\beta^{LRP}/\nu^{LRp}$ are reported in Table~\ref{LRptable}. We observed that, by using the Ising distribution, the Monte-Carlo measures on the LRp point are more precise than the corresponding ones for the fGFF. In particular we could probe with good precisions the physics of the LRp point for small values of $a$, the regime where the fGFF methods are more difficult to implement, see \cite{Javerzat_2020}. 

For $q=1$, we observed that, for $a<3/2$, the LRp is attractive while for $a\geq 3/2$ the system is described by the Bernoulli critical point Bp. For $q=2,3$ we establish the existence of the LR point, which is a long range critical point at finite disorder. We observed that the LR point and the LRp point exchange stability at a certain value $a^{*}(q)$ which depend on $q$. 

The above results are supported by the study of other observables than the FK fractal dimension. For $q=2$ we measured the multi-fractal behaviour of the spin-spin correlation, see Table~\ref{eta}. The results are consistent with the theoretical findings of \cite{dudka2016critical} in their region of validity, especially for $1.5<a<2$. For lower values of $a$, where theoretical predicitons are missing, our results rule out the occurence of a softening of the transition and support instead the fact that the system is driven to the LRp point. For $q=3$ we mesure the thermal $\nu^{LR}$ exponent by measuring the wrapping probability of the FK clusters. The results are shown in Fig.~(\ref{FigQ3_nu}) and are consistent, for a wide range of values of $a$, with the theoretical prediction of Eq.~(\ref{nulr}).  

We verified that the above mentioned results, in particular the form of the phase diagram, are valid for different disorder distributions. On the other hand, we observed some universal effects of the higher cumulants at the LRp point. By focusing  on the $a=1/4$ value, we showed that the exponent $\beta^{LRp}$ is different between a Gaussian and a non-Gaussian distribution, see Table~\ref{LRptable2}. Finally we pointed out that clear effects of the higher cumulants are expected for the universal finite size effects of the FK connectivities, see Eq.(\ref{p12}).

\section{Acknowledgments}

We are grateful to Nina Javerzat with whom this project started. We thank Pierre Le Doussal for useful discussions.

\appendix

\section{Disorder distributions}

\subsection{$n$-replicated Ising model}
\label{nising}
In most of the simulations presented here we chose a particular distributions for disorder. Let us consider  $n$ independent critical Ising models with spins $\sigma^{(a)}\in \{0,1\}$, $a=\{1,\cdots,n\}$. The variables $\sigma_i$ in Eq.~(\ref{bimodal}) are given by:
\begin{equation}
\sigma_i= \prod_{a=1}\sigma^{(a)}_i,
\end{equation} 
and their statistical properties are determined by the correlation functions of the the $n$ independent Ising models.  For the first two moments, we have:
\begin{equation}
\mathbb{E}\left[\sigma_i\right]=0, \quad \mathbb{E}\left[\sigma_i\;\sigma_j\right]\sim |i-j|^{\displaystyle -\frac{n}{4}} \text{ for ${|i-j| \gg 1}$},
\end{equation}
where we used the fact that the critical Ising spin-spin correlation function $\left<\sigma_{i}^{(a)}\sigma_{i}^{(a)}\right>\sim |i-j|^{-1/4}$.
Comparing with Eq.~(\ref{defa}), we have:
\begin{equation}
\label{avsn}
a=\frac{n}{4}.
\end{equation}
The convenience of this choice is that the $n$-Ising models are very simple, and fast, to be simulated. The price to pay is that we cannot vary $a$ continuously. 
The above distribution is not  Gaussian because the critical Ising point is not a Gaussian point. 

\subsection{Fractional Gaussian free fields}
\label{fgff}
The other one is related to the percolation point of the level sets of a fractional Gaussian random field with negative Hurst exponent $H=-a/2$.

Let us, now, consider an $L\times L$ square lattice, whose vertices $i$  have coordinates $i=(i_h, i_v)$ and a set of random variables $\phi_i$ that defines a discrete fractional Gaussian free field: 
\begin{equation}
\phi_i = \begin{cases}
\displaystyle \sum_{\displaystyle \substack{(l,m)\neq (0,0)}}^{L}  \displaystyle \frac{\displaystyle c_{l,m}}{S_{l,m}}\;\exp \left(\displaystyle\frac{2\pi}{L}(l\;i_h+m\; i_v)\right), \text{ if $(l,m) \neq (0,0)$} \\ 

c_{0,0}, \text{ if $(l,m)=(0,0)$}  
\end{cases}
\end{equation}

where $S_{l,m}= \left|  \displaystyle 2\cos\left(\frac{2\pi l}{L}\right)+\displaystyle 2\cos\left(\frac{2\pi m}{L}\right)-4\right|^{\displaystyle \frac{1-a/2}{2}}$
and the $c_{l,m}$ are independent random Gaussian variables, i.e. $c_{l.m}\in \mathcal{N}(0,1)$. Using the properties of the Fourier transform, the covariance $\mathbb{E}[\phi_i\phi_j]$ has the following asymptotic limit:
\begin{equation}
\mathbb{E}[\phi_i\;\phi_j]\sim |i-j|^{\displaystyle -a} \text{ for ${|i-j| \gg 1}$}.
\end{equation}

The $\sigma_i$ variables are defined by:
\begin{equation}
\sigma_i= \begin{cases}
0, \quad \text{if}\; \phi_i>0 \\
1 \quad \text{if}\; \phi_i<0, 
\end{cases}
\end{equation}
The fact that $\mathbb{E}[\sigma_i]=0$ implies the level $0$.
One can notice that the clusters built with the $\{\sigma_i\}$ variables are not at the percolation treshold. Indeed, this is spin-percolation physics and it is known \cite{Javerzat_2020} that it has a critical level greater than $0$.
These $\{\sigma_i\}$ variables are only needed to built the bond distribution as explained in Section~ \ref{numsetup}.
One can show, instead, that the FK-cluster built on these bonds, are at the percolation transition.
Thus, one has:
\begin{equation}
\mathbb{E}[\sigma_i\;\sigma_j]\sim |i-j|^{\displaystyle -a}\text{ for ${|i-j| \gg 1}$},
\end{equation} 
and a long-range correlated disorder.

\bibliographystyle{ieeetr}
\bibliography{biblio}

\end{document}